# The Golden Meteorite Fall: Fireball Trajectory, Orbit and Meteorite Characterization

P.G. Brown[1,2*] 0000-0001-6130-7039, P.J.A. McCausland[2,3] 0000-0003-3030-7524, A.R. Hildebrand[4] 0000-0001-9926-2059, L.T.J. Hanton[4] 0000-0001-5497-4554, L.M. Eckart[5] 0000-0002-8914-3264, H. Busemann[5] 0000-0002-0867-6908, D. Krietsch[5] 0000-0002-4289-1732, C. Maden[5] 0000-0002-6644-9535, K. Welten[6] 0000-0001-8577-6753, M. W. Caffee[7] 0000-0002-6846-8967, M. Laubenstein[8] 0000-0001-5390-4343, D. Vida[1,2] 0000-0003-4166-8704, F. Ciceri[4] 0000-0001-5497-4554, E. Silber[2,3] 0000-0003-4778-1409, C.D.K. Herd[9] 0000-0001-5210-4002, P. Hill[9] 0000-0002-6739-4169, H. Devillepoix[10,11] 0000-0001-9226-1870, Eleanor K. Sansom[10,11] 0000-0003-2702-673X, Martin Cupák[12, 10, 11] 0000-0003-2193-0867, Seamus Anderson[10] 0000-0002-8914-3264, R.L. Flemming[2,3], A.J. Nelson[13] 0000-0001-6398-7314, M. Mazur[1,2], D.E. Moser[14], W.J. Cooke[15], D. Hladiuk[16], Barbara Malečić[17] 0000-0002-7719-1146, Maja Telišman Prtenjak[17] 0000-0002-4941-8278, R. Nowell[18]

(The Golden Meteorite Consortium)

**\*Corresponding author email: pbrown@uwo.ca**

[1]Department of Physics and Astronomy, University of Western Ontario, London, Ontario, N6A 3K7, Canada

[2]Institute for Earth and Space Exploration, University of Western Ontario, London, Ontario, N6A 5B7, Canada

[3]Department of Earth Sciences, University of Western Ontario, London, Ontario, N6A 3K7, Canada

[4]Department of Geoscience, University of Calgary, Calgary, AB, Canada T2N 1N4

[5]Institute of Geochemistry and Petrology, ETH zürich, Clausiusstrasse 25, 8092 Zurich, Switzerland

[6]Space Science Laboratory, University of California Berkeley, Berkeley, California 94720, USA

[7]PRIME Laboratory, Department of Physics and Astronomy, Purdue University, West Lafayette, IN 47907, USA

[8]Laboratori Nazionali del Gran Sasso, Istituto Nazionale di Fisica Nucleare, Via G. Acitelli 22, I-67100 Assergi, Italy

[9]Dept. of Earth and Atmospheric Sciences, 1-26 Earth Sciences Building, University of Alberta, Edmonton, Alberta, T6G 2EG, Canada

[10]Space Science and Technology Centre, School of Earth and Planetary Sciences, Curtin University, GPO Box U1987, Perth, WA, 6845, Australia

[11]International Centre for Radio Astronomy Research, Curtin University, GPO Box U1987, Perth WA 6845, Australia

[12]Curtin Institute for Computation, Curtin University, GPO Box U1987, Perth, WA 6845, Australia

[13]Department of Anthropology, Western University, London, Ontario, N6A 5C2, Canada

[14]Jacobs/Space Exploration Group, EV44/Meteoroid Environment Office, NASA Marshall Space Flight Center, Huntsville, AL 35812, USA

[15]NASA Meteoroid Environment Office, Marshall Space Flight Center, Huntsville, Alabama 35812, USA

[16]RASC Calgary Centre, Calgary, Alberta, Canada

[17]University of Zagreb, Faculty of Science, Department of Geophysics, Horvatovac 95, 10000 Zagreb, Croatia

[18]College of the Rockies, Box 8500, Cranbrook, British Columbia, V1C 5L7, Canada

## Abstract

The Golden (British Columbia, Canada) meteorite fall occurred on Oct 4, 2021 at 0534 UT with the first recovered fragment (1.3 kg) landing on an occupied bed. The associated fireball was recorded by numerous cameras permitting reconstruction of its trajectory and orbit. The fireball entered the atmosphere at a 54º angle from the horizontal at a speed of 18 km/s. The fireball reached a peak brightness of -14, having first become luminous at a height of >84 km and ending at 18 km altitude. Analysis of the infrasonic record of the bolide produced an estimated mass of $78^{+157}_{-65}$ kg while modelling of the fireball light curve suggests an initial mass near 70 kg. The fireball experienced a major flare near 31 km altitude where more than half its mass was lost in the form of dust and gram-sized fragments under a dynamic pressure of 3.3 MPa. The strength and fragmentation behavior of the fireball were similar to those reported for other meteorite-producing fireballs (Borovička et al. 2020). Seven days after the fireball occurred an additional 0.9 kg fragment was recovered during the second day of dedicated searching guided by initial trajectory and dark flight calculations. Additional searching in the fall and spring of 2021/2022 located no additional fragments.  The meteorite is an unbrecciated, low-shock (S2) ordinary chondrite of intermediate composition, typed as an L/LL5 with a grain density of ~3530 kgm$^{-3}$, an average bulk density of 3150 kgm$^{-3}$ and calculated porosity of ~10%. From noble gas measurements the cosmic ray exposure age is 25±4 Ma while gas retention ages are all >2 Ga. Short-lived radionuclides and noble gas measurements of the pre-atmospheric size overlap with estimates from infrasound and lightcurve modelling producing a preferred pre-atmospheric mass of 70-200 kg. The orbit of Golden has a high inclination (23.5º) and is consistent with delivery from the inner main belt. The highest probability (60%) of an origin is from the Hungaria group. We propose that Golden may originate among the background S-type asteroids found interspersed in the Hungaria region. The current collection of 18 L - LL – chondrite orbits shows a strong preference for origins in the inner main belt, suggesting multiple parent bodies may be required to explain the diversity in CRE ages and shock states.

# Introduction

On Oct 3, 2021 at 2334 local time, a bright fireball was widely reported in the border area between the provinces of British Columbia and Alberta, near the town of Golden, British Columbia, Canada. A 1.3 kilogram L/LL5 chondrite penetrated the roof and landed on the bed of Ruth Hamilton in Golden, BC while she was sleeping. The accompanying fireball was recorded by several cameras in the region permitting reconstruction of the trajectory and pre-impact orbit of the meteorite. With an instrumentally measured orbit and sample available for study, Golden represents another clue as to the ultimate parentage for L/LL chondrites.

The identification of the asteroidal source bodies of the ordinary chondrites (OC) is a major unsolved problem in contemporary planetary science. One approach to deciphering OC origins is to measure their pre-impact orbits. The orbits for known meteorite falls provide a statistical constraint on the immediate escape region (either a mean motion or secular resonance) in the main belt from which meteorites of various classes have recently (on the order of Ma) emerged (Granvik and Brown, 2018). Orbital information for specific meteorite falls, when combined with meteorite data including spectral matches to near-Earth asteroids (DeMeo and Carry, 2014; DeMeo et. al. 2022) and dynamical models of the delivery of near-Earth asteroids/meteorites from the main belt to Earth (Granvik et al., 2018; Binzel et al., 2019), can ultimately constrain the location of OC parent bodies.

As of early 2023, more than 40 meteorite falls have published instrumentally measured orbits[1]. Including fireball producing meteorites with as-yet unpublished instrumental records, the tabulation of well-determined meteorite orbits is in excess of 50. L-chondrites are the most common class of meteorite falls (38% cf. Greenwood et al., 2020) and among published meteorite orbits where multi-station recordings were available to estimate orbits there are eleven from this class, roughly as expected based on the 38% fraction. Only three LL chondrite falls have known orbits, namely Chelyabinsk (Borovička et al., 2013), Stubenberg (Spurny et al., 2016) and Dishchii'bikoh (Jenniskens et al., 2020). At least one instrumental fall has a mixed classification (Dingle Dell L/LL6 (Anderson et al., 2023) and Innisfree may also be in this category (Granvik and Brown., 2018)).

---

[1] https://www.meteoriteorbits.info/

Pre-atmospheric orbits for L-chondrites provide useful constraints in defining the final stage of delivery for this most abundant of the OC groups to the Earth from the main belt. The precise source region and parent asteroid family which produce the L-chondrites, however, remains controversial.

Several unique characteristics of the L-chondrites factor heavily in constraining the original L-chondrite parent body (LCPB). Foremost among these is a peak in the K-Ar degassing age for the majority of L-chondrites at 470 Ma (cf. Swindle et al., 2014). This strongly suggests a common original origin for most L-chondrites associated with a major impact event in the main belt at this epoch. This collisional formation age is supported by an enhanced fall abundance of L-chondrites in Mid Ordovician time (Schmitz et al., 2001). A stratigraphically correlated range in cosmic-ray exposure (CRE) ages among these fossilized L-chondrites (Heck et al., 2004) suggests quick delivery of large amounts of L-chondrite material immediately following an asteroid family breakup. The high number of shocked L-chondrites and the slow inferred cooling rate for some of the degassed population is also suggestive of an origin in a large (>100 km) parent body (Nesvorný et al., 2009). The broad distribution of cosmic-ray exposure ages (Herzog and Caffee, 2014) among contemporary L-chondrites (with hints of peaks at 5 and 30 Ma) is consistent with transport of material from this earlier asteroid family breakup through a series of ongoing collisional cascades in the main belt together via Yarkovsky drift to efficient resonance escape "hatches" some distance from the original breakup location.

Based on these constraints, Nesvorný et al. (2009) suggested the Gefion family as a likely asteroidal source for the Ordovician (Schmitz et al., 2001) and contemporary L-chondrites. In their scenario, the 5:2 mean motion resonance (MMR) with Jupiter is the most accessible resonance for delivery of material to Earth immediately after family formation due to its proximity to the Gefion family and would be the escape region producing the ancient L-chondrite shower. However, contemporary L-chondrites could be delivered from more interior resonances such as the 3:1 after spending significant time undergoing Yarkovsky drift; they might even originate from another family such as Ino (Meier et al., 2017). This hypothesized link with Gefion has been called into question by recent work (eg. McGraw et al., 2018) who find that Gefion family members do not have reflectance spectra compatible with L-chondrites as well as others (Jenniskens et al., 2019; Devillepoix et al., 2022) who note that some L-chondrite orbits are incompatible with a Gefion source.

Additional early suggestions that L-chondrites could come from the Flora family (e.g., Nesvorný et al., 2007), have also been subsequently revised to suggest this family is more likely the source of the LL-chondrites based on spectral similarity and delivery efficiency of near-Earth asteroids (NEAs) and meteorites to the Earth from the nearby $\nu_6$ resonance (Vernazza et al., 2008; Dunn et al., 2013). Curiously, despite the fact that the debiased sample of LL chondrite-like NEAs are more than half of all NEAs (DeMeo et al., 2022) they represent only 9% of all meteorite falls, suggesting delivery differences with size from the main belt, a likely hall-mark of differing Yarkovsky drifts (Vernazza et al., 2008).

It is clear that while dynamical arguments and spectral linkages are insightful approaches in linking asteroid types with meteorites, more information is needed. In this respect, a large number of meteorite falls with well-determined orbits can aid in reducing the number of degenerate asteroid family – meteorite type links.

Here we describe the fall, recovery and orbit reconstruction of an unbrecciated weathering state 0 (W0) and shock state 2 (S2) L/LL5 meteorite which landed on the bed of a resident of the town of Golden, British Columbia, Canada. Together with its physical/chemical properties, we place it in context with the current suggested parent main belt asteroid families for the L and LL-chondrites.

## Circumstances of the Fireball and Associated Optical Instrumental Records

The fireball which produced the Golden meteorite first appeared at 05:33:43 UTC (22:33:43 MST local) on Oct 4, 2021 (Oct 3 local time) near the border of the Canadian provinces of Alberta (AB) and British Columbia (BC) at >84 km altitude. Weather conditions at the time were mixed with some cloud and rain in the area of the fall (Fig. 1). The ground path parallels the border, occurring roughly 50 km inside BC with an endpoint just to the west of the town of Golden (Fig. 2).

The fireball was recorded in AB and BC by 30 eyewitnesses submitting reports online to the American Meteor Society (Perlein, 2021). Among these were a handful near Golden, BC who reported sonic booms. One eyewitness, Ruth Hamilton, also living in Golden reported on the AMS site that the fireball "hit and entered my house". This proved to be the initial report of a meteorite which had penetrated Hamilton's roof and landed on her bed as described in detail in the next section. Subsequent field work located additional eyewitnesses in the immediate fall area in the course of meteorite recovery efforts.

A number of optical instrumental recordings were made of the associated fireball. These included several dashcam/security records and at least four camera/photographic observations from dedicated fireball cameras. Table 1 provides an overview of all available video/photographic recordings for which astrometric and/or photometric calibrations were possible.

The first photographic record to be widely circulated on social media of the fireball was by Hao Qin, who happened to be taking a sequence of nighttime photographs over Lake Louise, AB at the time of the event. During one 15 second-long Nikon DSLR exposure in his sequence he captured the early and mid-portion of the fireball; the raw image had apparent saturation of the sky near the fireball trajectory, but his image processing produced a result portraying the fireball along with many nearby stars enabling positional calibration (Fig. 3). The later part of the fireball is expressed after processing as a low contrast streak and it is not clear if the shutter closed before the fireball ended. When contacted, Hao Qin helpfully provided a high-quality version of this processed image which allowed an initial precise trajectory solution by Oct. 6 (along with other serendipitous images).

The all-sky fireball camera operated by Don Hladiuk near Calgary, AB as part of the Global Meteor Network (Vida et al., 2021), station CA000J, also detected the fireball and produced a record almost immediately. While saturated in mid-late flight, the early portion of the trajectory was useable for astrometric reduction.

One of us (ARH) through active solicitation of facilities with security cameras close to the fireball endpoint was able to secure a variety of IP (Internet Protocol) security camera footage including from cameras directly imaging the fireball from Sunshine Village Ski Resort in AB. Though several frames are heavily saturated, the beginning and end of luminous flight were well captured and the camera was sufficiently sensitive to allow direct calibration from stars visible on each frame as shown in Fig. 4. Note also the presence of two discernable fragments in the final frame.

From these the two most proximal stations, an initial solution located the fireball endpoint just to the west of the town of Golden, BC (Fig. 2). Dark flight calculation outlined the meteorite fall zone and a public media release which indicated that residents of Golden should be alert for the possible presence of meteorites. It also guided initial in situ recovery searches by ARH and LTJH which led to the recovery of a second fragment as described in the next section.

Additional fireball imagery from a Spalding Allsky Camera network[2] station near Cranbrook, BC provided video from the South-Western side of the fireball track allowing additional astrometric measurements. A security camera video from Delacour, AB captured the full fireball flight (Fig. 5) which provided excellent relative photometry, but astrometric calibration was not possible for logistical reasons. A dashcam in a moving car captured the entirety of the fireball from Calgary and proved useful for relative photometry.

---

[2] http://goskysentinel.com/node/node28

Finally, two dedicated fireball cameras of the Global Fireball Observatory (GFO) project (Devillepoix et al., 2020) at Mattheis Ranch (name of landowner and camera host) and Vermilon (town nearest camera station) operating as part of the Meteorite Observation and Recovery Project (MORP) 2.0 network in Alberta captured the fireball at extreme ranges of 480 and 375 km (Fig. 6) respectively. Though at low altitudes, the high resolution of these cameras permitted good astrometry and (for Vermilion) reasonable photometry.

For all cameras, the fireball astrometric calibration was performed using the SkyFit2 package (Vida et al., 2021). Different distortion models were tried for each camera and the model showing the smallest fit errors overall was chosen for reduction. For the GFO cameras, the best model was found with a radial distortion fit using a $7^{th}$ order polynomial keeping odd terms only. The root mean square (RMS) error in stellar astrometry was found to be 1.1 arcmins for Mattheis Ranch and 0.9 arcmins for Vermilion. For Sunshine the best model fit was found using a $3^{rd}$ order polynomial combined with $5^{th}$ order radial term which produced RMS residuals of 0.8 arcmins. For Cranbrook, a $5^{th}$ order radial with odd terms produced RMS fits of 7.2 arcmin and for the GMN CA000J camera a model with $7^{th}$ order odd polynomials produced a 2.8 arcmin RMS residual.

The initial trajectory solution using all astrometry from all cameras with the approach described in Vida et al. (2020) (which uses timing and astrometry simultaneously) produced a reasonable global fit. Note that for the Lake Louise image, the brightness as a function of time was used to locate brightness "dips" in the early part of the lightcurve which could then be tied to common features from cameras which had timing information. This produced four pseudo-time points from Lake Louise. These data were not used in the solution, but the look angles were mapped onto the master solution. Figure 7 shows the resulting spatial residuals from all stations while Fig. 8 shows the fireball ground path relative to all stations. As Lake Louise was the closest station, the fact that the spatial residuals are all less than 100 m from the main trajectory is reassuring. The main outlier here is the Sunshine data which has a systematic westward shift compared to the overall best solution. This may be partly attributed to the poor look angle or a change in the frame rate which affects trajectory solutions with this technique as good relative timing is required. As this difference is most pronounced at the end of the trajectory, another possibility is that our differential refraction model is not fully correcting the elevation angles for more distant stations. The University of Calgary group has noticed similar positional discrepancies for one other fireball trajectory (for the end point) with differences between proximal image reductions vs. those including distant images (taken at sites located to the east over the prairies); the positional discrepancies significantly exceed what is allowable from variation due to measuring precision indicating at least one perturbing physical effect.

Removing Sunshine from the best-fit trajectory does not significantly alter the overall radiant (less than 0.5 deg difference). The resulting best-fit trajectory is summarized in Table 2 and the orbit is given in Table 3 while the orbital covariance matrix is shown in Table 4. The initial speed is estimated based on the average speed in the first portion of the trajectory using the iterative method described in Vida et al. (2020).

Finally, we note that a geometric solution using only Lake Louise and Sunshine produces a radiant difference from our nominal best fit solution of close to 1.5 degrees. Such a systematic shift would produce equivalent potential systematic errors about four times larger than the random errors given in Table 3. However, given the consistency in lags and spatial residuals between all stations (except Sunshine) we view this as an extreme upper limit.

## Details of Meteorite Recovery/Fall Circumstances

Two meteorites of kilogram mass from the Golden fireball have been recovered as of March 2023.

The first, hereafter also referred to as the Hamilton fragment, penetrated a roof and landed on the bed of Ruth Hamilton in the town of Golden, at location 51°17'43.3"N, 116°57'29.6"W. Her description of the event shortly after 23:30 on Oct 3, 2021 local time is as follows:

*"I was sound asleep in bed at the time of the meteorite fall. The only warning I had about what was to happen was when my dog started barking and woke me up shortly before the impact, but then I went back to sleep. Next, I heard what sounded like an explosion and debris of plaster and shingles all over my face and bed. I thought the window had broken maybe because a tree nearby had fallen. I then noticed a strong burning hot smell, like something hot on the stove. I jumped out of bed and ran across the room to turn on the light switch, saw the clock read 11:35 pm, and then noticed a small hole in the ceiling and drywall hanging just over my bed. I immediately called 911 and told them there was a hole in the ceiling over my bed. The operator then asked me some questions and asked me to go back into the bedroom. When I went in I flipped over a pillow and said "oh my gosh there is a rock in my bed". The rock had slipped in between the two pillows I was sleeping on without my noticing. An officer showed up within just a few minutes of my call. He said he had been a few blocks away and had heard a loud explosion just 5 minutes before he arrived at my door. The officer came inside and pushed the rock on the bed around with his flashlight to try and figure out what it was. There had been some highway blasting occurring nearby over the last few weeks and he initially thought this might be related to that work. He went to his car and phoned the highway crew and they indicated no blasting had occurred but the workers had seen the fireball. The officer then came in and suggested that it was a meteorite from the fireball the workers had witnessed and then he left. I didn't sleep the rest of the night and just stayed awake in the living room. "*

The corresponding written report by the Royal Canadian Mounted Police (RCMP) includes:

*"On 2021-10-03 at 23:50 hrs (MST) Golden RCMP received call from Ruth Hamilton advising she heard a large bang and then looked up to find a hole in her ceiling and a fist sized rock on her bed. Hamilton advised the rock gave off a burnt metallic smell and appears it was the rock the created the hole down through her ceiling.*

*Cst Fraser had heard a large boom about 10-15 minutes prior to this call but was unable to see the direction it was coming from. Cst Fraser attended Hamilton's residence and checked the scene and noted the rock would have been coming at*

*high speed from the sky to have punctured a metal roof, through an attic down onto Hamilton's bed. Cst Fraser checked with Kicking Horse Canyon Project to see if they were doing any blasting tonight and was advised no but that construction workers witnessed a large bright ball in the sky burning past them and they seen the object explode sending projectiles in all directions near Golden BC. Cst Fraser determined that this rock is probably a meteorite and given the speed it was falling would be enough to crash through a roof. Cst Fraser followed up with Hamilton and left meteor with her and provided file number for insurance. Cst Fraser also took photos of scene for hardcopy file. No further action required."*

The RCMP report also noted that the call from Hamilton was received at 22:50:21 MST, that Cst Fraser was on the scene at 22:57:01 and that his onsite investigation concluded at 23:20:49 MST.

Figure 9a shows the 1.27 kg Hamilton meteorite in the orientation it came to rest on the bed and ceiling damage in the bedroom, while Figure 9b shows the ingress hole on the tin roof from the meteorite. Figure 9c shows this fragment in closeup. The meteorite is roughly brick shape with dimensions of 11x7x6 cm.

Following this initial recovery, ground searches initially by two of us (ARH, LTJH) in the predicted fall zone produced recovery on Oct 10, 2021 of a 0.92 kg mass, hereafter referred to as the "Calgary" fragment (Fig. 9d), on a roadway 1.32 km NNW of the main mass fall location. Further efforts at meteorite recovery in the autumn of 2021 and spring of 2022 were unsuccessful. The street and alley grid (oriented to the cardinal points) of central Golden were all searched by bicycle during the first week after the fall (the street and alley surfaces equal ~10% of an ~1.6 × 0.8 km area but lawns and yards were viewed as well); the Hamilton piece was recovered within this area and the substantial coverage by buildings (~20% by area), lawns and paved areas frequented by Golden residents constrain the possibility that any other similarly sized meteorites fell in this part of Golden. Residents' interest in recovering meteorites was shown by the number of suspected meteorite samples shown to searchers. As well, baseball/sports fields, school yards, gravel pits, airport tarmac and other favorable areas distributed within the projected strewn field were searched on foot. Approximately 40 km of additional streets and roads were searched by vehicle within the strewn field yielding the one road side specimen (0.92 kg "Calgary fragment"). In the spring of 2022 ~10 hectares of cut hay fields near the Golden Donald Upper Road – Barber Road intersection were searched with a magnetic device in good recovery conditions (based upon past experience), but no meteorites were found. However, the Golden meteorite being an L/LL type, has only ~3% Fe-Ni metal grains so its attraction to magnets is likely reduced by roughly an order of magnitude relative to that of the H chondrites previously recovered with this device in other strewn fields (hand tests with a magnet confirmed significantly lower attraction). This makes interpretation of the negative search result less clear.

**Meteorite Darkflight Modelling**

From the trajectory found earlier, we model the darkflight of fragments from the end of the fireball to compare with actual recovery locations. This provides an independent check on the accuracy of our trajectory solution and also an indication of where additional material may have fallen.

For our darkflight model, we use the Western Meteor Group darkflight Monte Carlo (MC) code (Shaddad et al., 2010; Brown et al., 2011). In this model, a nominal fragment with user-defined mass/area ratio is followed to the ground under the action of wind using the basic approach described by Ceplecha et al. (1987). The uncertainty in this ground location is then defined using a series of realizations about this nominal solution incorporating uncertainties in the radiant and fireball end point.

For the Golden fireball, we released all fragments at the end location given in Table 2 using the computed state vector appropriate to the fireball endpoint. We used the wind and temperatures derived from the Weather Research and Forecasting Model (WRF) (Skamarock et al., 2019) at the time and location of the fireball endpoint. We further assumed a velocity at release of 4 km/s, consistent with the observed terminal speed and following Ceplecha et al. (1987) a bulk density of 3700 kgm$^{-3}$ and a spherical shape. We note that for a fixed shape, the darkflight locations are all equal for the same mass/cross-sectional area. Hence, assuming a lower bulk density (such as 3200 kgm$^{-3}$ as measured) scales the equivalent mass to be 1.3 times higher.

Figure 10 shows the ground footprint from 200 MC runs for each mass interval from [5 g, 5 kg] using uncertainties from the computed radiant and ground location. The larger 1.3 kg mass (which hit Ruth Hamilton's bed) is at the extreme downrange portion of the kilogram-sized footprint while the 0.9 kg fragment is embedded in the 0.5 kg footprint, about 400 m from the nearest darkflight MC kilogram-class fragment. Note that the relative agreement is even better if we adopt a bulk density of 3200 kgm$^{-3}$ instead of 3700 kgm$^{-3}$. The MC spreads are conservative as we have assumed no lateral velocity spreads, though these are commonly observed for fireballs (eg. Borovička and Kalenda, 2003), nor explored differences in fragment shape. Assuming brick-like shapes or hemispherical shapes will shift the footprint toward the endpoint. Given these uncertainties, the agreement between recovered locations and that predicted from our darkflight model is quite good using our nominal trajectory solution and provides confidence in the computed fireball path.

Our darkflight modelling also predicts that a 1.3 kg fragment released at the endpoint would reach the ground after about 130 sec of darkflight, namely at 11:36:00 pm local time. This timing is in excellent agreement with Ruth Hamilton's observation that her clock read 11:35 pm when she was suddenly awakened by the meteorite penetrating her roof.

We note that any smaller fragments (which may be more numerous) would be blown by the lower atmosphere winds farther East and into forested/mountainous terrain making recovery difficult.

**Fireball Lightcurve**

Only three recordings proved useable for photometric measurements. From the Vermilion GFO station, while the fireball was low on the horizon the complete middle (including the brightest) portion of the lightcurve is visible. Because of the range, the peak apparent brightness is fainter than -9 mag, just at or slightly below the saturation level for GFO cameras. The fireball was only a few degrees above the local horizon as seen from Vermilion making photometric correction challenging. We used the extinction model of Green (1992) as implemented in Vida et al. (2021) and adjusted the extinction parameters to minimize photometric residuals for stars near the elevation and azimuth of the fireball. We note that adjusting the extinction to better fit high elevation stars produced a 0.8 magnitude decrease in brightness compared to our best estimate using low-elevation stars, indicating a high gradient in the extinction model at low elevations.

The complete lightcurve was also visible on the Calgary Dashcam and the Delacour security video (see Table 1) and neither was saturated. The relative brightness from these two cameras was computed and then scaled to the peak Vermilion absolute brightness of -13.9 to produce the best estimate for the full fireball lightcurve as shown in Figure 11. As the absolute timing and sampling were different for all three cameras, common features (such as the bright flare near 30 km altitude) were used to align the timebase. The photometry procedure was performed using the Skyfit2 program as described in Vida et al. (2021).

The resulting peak absolute magnitude is near -14. An independent check on this value was made by searching for the detection of the fireball in Geostationary Lightning Mapper (GLM) data from both GOES-16 and 17 satellites. GLM is known to be able to detect slow fireballs (speeds < 20-30 km/s) with peak magnitudes of order -14 (Jenniskens et al., 2018) or brighter. No detections were correlated either spatially or temporally with the timing of the Golden fireball, consistent with our ground-based peak magnitude of -14. The location of the fireball was near the edge of the GLM field of view for GLM from both satellites (Smith et al., 2021) so we expect the actual threshold limit to be brighter than -14. As an empirical check, we note that Jenniskens et al. (2018) described the Crawford Bay meteorite-producing fireball of Sep 5, 2017, which was detected by GLM and had a very similar nadir angle to the Golden fireball relative to the GOES satellites. They found that for the Crawford Bay fireball, the ground-calibrated threshold for GLM was -15.3, consistent with our non-detection of Golden from the GOES-GLM instrument.

As an additional observational check, eyewitnesses described the mountains brightening more than from a Full Moon and security cameras from the Kicking Horse Mountain Resort (located close to directly beneath the fireball end) recorded a strongly distinct shadow from the brightest flare. One family watching television inside their home noticed the bright flashing illumination outside their windows (and shortly afterward heard the associated boom arrive – the boom was also heard by construction workers who were outside).

The light curve shows a main flare near 30 km altitude and a smaller flare at 34 km. A broad plateau in brightness noticeable on all cameras between 50 – 65 km hints at small amounts of higher altitude fragmentation. Two of the three cameras show a suggestion of a small flare near 45 km but this is less certain. The earliest reliable luminosity was detected above 84 km altitude and the final two fragments were still visible at 18.5 km.

**Ablation Modelling**

Using the observed lightcurve and apparent lags (difference between the observed along path position compared to a constant velocity model) as input data from each station, an ablation modelling fit was performed to establish the likely initial mass. Here we use the semi-empirical model of Borovička et al. (2013) with the updated luminous efficiency dependence on mass/velocity presented in Borovička et al. (2020). We assume a bulk density of 3200 kgm$^{-3}$, and a product of the drag coefficient (Γ) and shape factor (A) of 1.0. The observed entry angle of 54 degrees is used and the initial speed adjusted to minimize the spread in the lags between stations. We assume an intrinsic ablation parameter of 0.002 s$^2$/km$^2$ and, following the approach of Borovička et al. (2020), perform a manual fit by releasing eroding fragments or fragments as appropriate to match the lightcurve. The grains released were assumed to follow a differential mass power law with index s=2.0 with an adjustable grain mass range and a variable erosion coefficient adjusted at each fragmentation point. The fraction of the total meteoroid mass released in any given fragmentation event was also a tunable parameter. As the dynamics of the fireball (except large lags near the end) were not observed with high precision, our main constraint is the lightcurve with time/height. The resulting fragmentation behaviour, mass released, grain distribution, etc. should be taken as representative only and not unique. The total initial mass to produce the adopted lightcurve is expected to be more secure, assuming our model of luminous efficiency is correct.

Table 5 summarizes a fragmentation scheme which was found to fit the lightcurve and observed lag simultaneously. Figure 12 presents the model lightcurve and lag compared to observations. The main features of this solution are:

a) Initial fragmentation of some 5% of the initial meteoroid mass around 65 km altitude as small (sub-gram) grains which produces the plateau in brightness between 65 – 50 km. Prior to this height the body followed single-body ablation.
b) Production of the main flare near 30 km by release of some 85% of the remaining meteoroid mass including fragments up to 100 g in mass. This occurred under 3.3 MPa of dynamic pressure.
c) Survival of two main fragments – one (from the main body) at the fireball end height of 0.9 kg and another eroding fragment released near 24 km of 6.9 kg. The latter continues to ablate and reaches a mass of 1.0 kg before beginning darkflight.
d) The initial mass from this model is 70 kg with a range of luminous efficiency of 3 – 6% over the height interval from 30 – 65 km.

The peak dynamic pressure experienced by the fireball was 3.5 MPa at 26 km, with the first fragmentation occurring at 0.04 MPa. For context, Borovička et al. (2020) showed that meteorite-producing fireballs typically fragment in two discrete dynamic pressure intervals: the first near 0.04 – 0.12 MPa and a second in the range of 0.9 – 5 MPa. In terms of apparent meteoroid strength, the Golden fireball is therefore unremarkable.

The measured ultimate compressive strengths of samples from the Calgary fragment are 32.5 to 43.9 MPa with a mean of 37.7 MPa (measured with a TestResources 313Q Universal Test Machine electromechanical press at the University of Calgary) (Ciceri et al. 2023). Thus the Golden meteorite fireball is typical in having fragmentation occur at dynamic pressures at least an order of magnitude less than the rock's compressive strength (e.g., Borovička et al., 2020).

The only somewhat unusual property of the Golden fireball was that most of the material was ablated during the second (later) fragmentation phase, which Borovička et al. (2020) found occurred in only a minority of their observed sample (e.g., Jesenice (L6) and Renchen (L5-6) falls).

## Fireball Infrasound Detection/Analysis

Infrasound from the Golden fireball was detected at two infrasound arrays operated by the Comprehensive Test Ban Treaty Organization (Le Pichon et al. 2019). Signals were found and analyzed following the procedure described in Ens et al. (2012) and Gi and Brown (2017) at the I56US and I10CA arrays located at ranges of 330 km and 1483 km, respectively. Propagation to I10CA was aided by a strong tailwind at stratospheric altitudes. At such ranges, the source energy is best determined using the observed period at maximum amplitude which is less affected by the cumulative effects of wind as compared to pressure amplitude (Silber and Brown, 2019). Table 6 summarizes the signal characteristics for each station while Figure 13 shows the signal as determined by locating the maximum in the time-windowed cross-correlation across all array elements. Note that the two strong wavetrains at I56US are likely due to multipath propagation, a common occurrence in infrasound signals (Le Pichon et al., 2019). For the period estimate at each station, four independent estimates of the period were used. One used a simple zero crossing estimate for period at maximum amplitude (ReVelle, 1997) while the second was computed by finding the maximum power spectral density of the signal window containing the bolide (Ens et al., 2012). Two additional independent estimates of the signal period were performed using the Progressive Multi-Channel Correlation array tool (Brachet et al., 2010). Specifically, arrival families for the Golden fireball acoustic signal were identified based on common backazimuths and the signal mean max period and pseudo period were found. Table 7 summarizes all signal period estimates.

The best source energy estimate for fireballs where multiple stations have period measurements was found by Ens et al. (2012) to be from a simple period average (see their Fig. 16). Using this multi-station period-energy relation we find a best source energy estimate of $0.003^{+0.006}_{-0.0025}$ kt TNT equivalent, where 1 T TNT is 4.184×$10^9$ J. For an entry speed of 17.9 km/s this corresponds to mass bounds of $78^{+157}_{-65}$ kg or equivalently for a measured bulk density of 3200 kgm$^{-3}$ a diameter range of 36±16 cm.

## Golden Meteorite Samples and Classification

## Golden Meteorite Sample Examination

Ruth Hamilton's main mass (Fig. 9a-c) was examined by one of us (MM) during an initial visit to Golden and was loaned to Western University for investigation and description. The Hamilton fragment is a complete, rounded brick-shaped individual entirely covered in fusion crust. One surface shows broad regmaglypts and a smaller, square-end surface with minor roll-over rims has thinner fusion crust through which large ~mm-size chondrules can be observed. Many surfaces of the Hamilton mass are modified with adhered roofing materials from having punctured the tin, asphalt, plywood and drywall roof during its arrival. The second Calgary fragment was found by ARH and LTJH and also has a blocky, rounded shape with 70% fusion crust and one significant broken surface revealing a light grey stony interior, bearing some well-defined dark chondrules (Fig. 9d). The fusion-crusted surfaces of the two individuals are similar in maturity and appearance, suggesting that they are from the same fall.

The complete 1.27 kg Hamilton fragment (roughly 11×7×6 cm) was imaged in three dimensions using an Artec Space Spider 3D Scanner at the Department of Anthropology at Western University, by placing the fragment on a rotating plate with distinctive marks for image registration during scanning. Scans were taken in different spin orientations to obtain images of all surfaces and aligned to produce a 3D model of the fragment with 0.1 mm resolution. Color imaging during the scan process allowed the final Hamilton fragment model to have appropriate surface color and albedo features, including metallic scratches and an adhered dark asphalt patch from its passage through the roof of the house (Figure 14).

The 0.92 kg Calgary fragment was color imaged from all sides in diffuse white light using a Nikon D850 at the University of Calgary's Library and Cultural Resources Digital Services; 75 focus-stacked images were assembled into a 3D model using Meshroom (Figure 14). The fragment was also scanned to construct a scaled 3 D model (of known volume) with a NextEngine Ultra HD 2020i laser scanner.

Micro-X-Ray computed tomography (µCT) imaging was done for the complete Hamilton fragment as well, using a Nikon Metris XTH-225ST micro-computed tomography system housed at the Museum of Ontario Archaeology in London, Ontario. Canada, equipped with a microfocus X-ray source and a reflecting tungsten target, operating at 225 kV, 147 µA with a 1.0 mm Sn filter at source to reduce beam hardening. The Hamilton fragment was mounted standing on its long axis, and 3141 projection images were taken during sample rotation through 360°. The capture software was X-Tek Inspect-X v 4.4, and the projections were reconstructed using X-Tek CT Pro v 4.4, producing a 3D reconstructed volume with 60 μm voxel-edge resolution (Fig. 14b). Using Dragonfly Pro 2022 2.0 software, these data enabled a bulk volume to be determined for the Hamilton fragment as well as the integrated % volume occupied by high X-ray absorbing phases metal and sulfides. The manipulation of the 3D reconstruction also permitted the investigation of internal textures and a projection of the type sample cut.

Bulk petrography is possible with the μCT volume of the Hamilton fragment, to the limit of its 60 μm voxel-edge resolution. The metal+sulfide portion of the entire main mass volume represented by thresholding the higher voxel intensity values is ~8%, intermediate between L and LL-chondrites (Brearley and Jones, 1998). Metal and sulfide occur as disseminated grains and occasionally as mm-wide, irregular aggregates of ~cm area extent (Fig. 15), confirmed in cut surfaces. Several metal-poor, X-ray absorbing, subrounded inclusions of up to ~cm size are observed in the Hamilton μCT volume, likely corresponding with fine sulfide-bearing dark inclusions. There is no evidence for brecciation in the μCT volume.

**Golden Sample Handling and Bulk Physical Properties**

The Hamilton mass was wire saw cut on one corner diagonal plane to produce a type specimen (hereafter referred to as the Golden type specimen) that served, unless otherwise noted, as the source for all further petrographic, mineral chemistry, X-ray diffraction, physical property, noble gas and cosmogenic nuclide analyses representing the original 1.27 kg Hamilton fragment. A 13.53 g portion of the Golden type specimen (Fig. 16) was selected for non-destructive gamma-ray spectroscopy (details below). The Calgary 0.92 kg fragment was cut to produce a sample for petrographic and mineral chemistry analyses, as well as a cut block for physical property measurements, and provided aliquot samples for parallel noble gas and cosmogenic nuclide analyses. Subsampling for the aliquot samples from both Golden fragments was done at >1 cm depth from the fusion crust, to avoid the fusion heat-affected zone in the meteorite near surface and terrestrial alteration (detailed aliquot sample methods and results reported below).

The meteorite bulk density determined for the complete 1270 g Hamilton fragment using the μCT volume is 3203±10 kgm$^{-3}$. The 13.53 g Golden type specimen was measured for grain (intrinsic) volume via He gas pycnometry using a Quantachrome Multipycnometer at Western University, giving a grain density of 3569±10 kgm$^{-3}$. The same specimen measured with a Sapphire Instruments SI-2 magnetic susceptibility meter produced a value of (SI units $10^{-9}$ m$^3$kg$^{-1}$) log χ = 4.72. The grain and bulk densities obtained from the type specimen result in a calculated porosity of 10.4%, which is visible as sub-mm pits in the cut faces of the meteorite (Fig. 16). A 28.28 g cut block of the Calgary fragment with dimensions measured by digital caliper returned slightly lower values (with similar instrumentation at the U of Calgary) than the Hamilton fragment, having grain density of ~3490±3 kgm$^{-3}$ (1σ uncertainty from eight repeated measurements), bulk density of ~3095 kgm$^{-3}$, with calculated porosity of 11.3%. These physical properties are consistent with both L and LL chondrite meteorite falls (Flynn et al., 2018), although the magnetic susceptibility is higher than expected for typical LL falls (Rochette et al., 2003). Magnetic isothermal remanence saturation experiments on two chip samples from the Hamilton fragment (253 mg and 120 mg) return S$_{300}$ values of 0.48 and 0.55, indicating the presence of coercively hard tetrataenite in the metal fraction, more consistent with LL-chondrites (Gattacecca et al., 2014).

## Golden L/LL Chondrite Mineralogy and Petrography

Cut surfaces of Golden reveal dark chondrules and disseminated mm-sized metal grains in a green-grey matrix, suggestive of an L chondrite based on the amount of metal content (Fig. 16). In situ 2D X-ray diffraction (2D XRD) analysis of distributed 300 μm crystalline spot locations in two Golden cut surfaces using a Bruker D8 Discover Micro X-ray Diffractometer (Flemming, 2007; detailed methods in Rupert et al., 2020) typically identify ferroan forsterite, enstatite and minor troilite phases. Short diffraction streak lengths are observed in olivine and pyroxene grains indicating crystal deformation consistent with shock stage S2 (Rupert et al., 2020). Using the unit cell refined for olivine from multiple diffraction peak positions (Round et al., 2010), the estimated olivine composition is ~$Fa_{24}$, most consistent with L-chondrites.

Optical microscopy of polished thin sections shows abundant large chondrules and chondrule fragments of POP, PO, RP and BO types, moderately delineated in a recrystallized fine to medium grained feldspathic matrix. Chondrule mesostasis is finely recrystallized as feldspar. In the Calgary specimen, additionally a single opaque chondrule occurred comprised of very fine-grained spinel set in feldspathic matrix with a recrystallized coarser rim (Appendix C). Accessory matrix phases are apatite and chromite. Olivine and pyroxene grains exhibit sharp to undulatory extinction, consistent with shock stage S2 (Stöffler et al., 2018; 2019). Porosity (~10 vol%) occurs primarily as irregular, 100 μm wide pits between chondrules. In reflected light, troilite is present as abundant (5 vol%) interstitial grains and also as small inclusions in some chondrules and decorating some chondrule rims. Fe-Ni metal is less abundant than the troilite, occurring as disseminated interstitial grains and as larger, anhedral grains and aggregates up to mm in size.

These petrographic features, particularly sulfide:metal ratio >1, are most consistent with an LL5 meteorite, albeit with somewhat high metal content. Weathering phases are absent.

Electron probe microanalysis (EPMA) was performed on several thin sections and a polished puck, using a JEOL JXA-8530F Field Emission electron microprobe at the Earth and Planetary Materials Analysis Laboratory at the University of Western Ontario, and aJEOL JXA-8200 electron microprobe at the UC Lab for Electron Microprobe Analysis at the University of Calgary. Operating conditions were an accelerating voltage of 15 kV, beam current of 20 nA and a beam spot size of 5 μm. Natural and synthetic standards were used for calibrations (Jarosewich, 2002). EPMA results are given in Appendix A.

Backscattered electron images and X-ray elemental maps for the polished puck are available in Appendix B. Maps for P and Ca in particular indicate that apatite and merrillite are widely distributed as 30 to 200 μm grains, and may be inhomogeneously distributed at a small enough volumetric sampling scale, of significance for noble gas interpretation below. Sulfide and metal distribution is indicated by S, Fe and Ni maps, and is consistent with the reflected light petrographic observations indicating that sulfide is more abundant than metal, although metal can occur as much larger aggregates of up to mm size in section, and as larger plate-like aggregates (Figs. 15, 16).

Matrix and chondrule olivine grains in Golden are strongly equilibrated, with mean composition $Fa_{26.3\pm0.2}$ (n=53). Ca-poor pyroxene is similarly equilibrated, as $Fs_{22.5\pm0.2}$ $Wo_{1.3\pm0.2}$ (n=49). The limited silicate compositional variation is consistent with type 5, but the compositions fall within overlapping ranges for L and LL-chondrites (Grossman and Rubin, 2006).

Golden is classified as an ordinary chondrite L/LL5 (S2) W0 (Gattacecca et al., 2023). EPMA analyses of olivine and Ca-poor pyroxene do not discriminate between L and LL groups for this meteorite. Other properties are also equivocal: grain density and bulk density values are consistent with both L and LL-chondrites (Flynn et al., 2018); observed sulfide:metal ratio >1 and high magnetic coercivity of the metal suggest that this is an LL chondrite. The bulk magnetic susceptibility of log $\chi$ = 4.72 is more consistent with an L chondrite. Golden does not appear to be a breccia, so there is little scope for explaining the apparent L versus LL contrasting characteristics as being due to nonrepresentative sampling bias.

**Golden fragment shape analysis**

With only two pieces recovered and a relatively small original mass the possibility that these two meteorites represent the last two fragments imaged at the fireball's end and may have been "mated" has been considered. The two pieces mostly have similar degrees of ablation smoothing on their faces. They also share one very flat surface – indeed these are the flattest surfaces on meteorites' that some of the authors have ever seen. Its presence suggests that the Golden meteoroid may have contained a healed joint surface to eventually be a preferential failure plane. One fusion surface on each individual fragment shows evidence ( e.g., anomalous brownish colour and/or spattered fusion crust ring) of minor late oriented flight. Both fragments' shapes are better approximated by cubes than spheres suggesting that drag coefficients of 0.8 – 1.0 are more appropriate in modelling dark flights (rather than ~0.5); the shapes' larger drag coefficients are also consistent with the development of oriented flight. The Hamilton fragment seems a better candidate as one of the later imaged fragments as its fusion crust is almost intact. In contrast, the Calgary fragment has one large broken surface and several small breaks possibly indicating that it hit other rocks in dark flight (after ablation ended) or otherwise fragmented. The breaks are thought not to be due to ground contact as extensive searching didn't reveal any broken pieces in the area, and the meteorite's surface scuffed from ground contact isn't broken. Dark flight modelling also suggests that the Calgary fragment was released from the main meteoroid higher along the trajectory. Given the lack of recovery prospects in much of the projected strewn field's area, statistically several additional fragments of similar size may be expected on the ground. Both the actual meteorites and 3D printed models of the meteorites were manipulated to a find possible fitted orientation. Figure 17 shows the apparent best fit although the contact surface matching process is obscured by some ablation and a late-stage partial broken surface. This possible mating orientation is compatible with cosmogenic nuclide analyses, which indicates that the two rock sampling positions are ~10 cm apart (see cosmogenic nuclide section).

## Golden Bulk Composition

In total two samples of the Golden meteorite, a 92 mg aliquot from the 1.27 kg main mass (labeled Hamilton), and a 52 mg aliquot from the 0.92 kg fragment (labeled Calgary) were selected for bulk analysis. We gently crushed both samples in agate mortars. For the Calgary sample, we used a bulk sample for radionuclide analysis, while for the Hamilton sample, we separated the magnetic (metal) fraction from the non-magnetic (stone) fraction and only used the non-magnetic fraction. The magnetic fraction of 4.5 mg of the Hamilton sample was ultrasonically agitated in 0.5% HCl to remove the attached troilite. After rinsing the metal several times with ultrapure water, once with ethanol, the metal fraction was dried and weighed, yielding 4.0 mg of relatively clean metal, corresponding to 4.5 wt% of the bulk sample. The metal sample was dissolved in dilute $HNO_3$, and an aliquot was taken for chemical analysis by ICP-OES, yielding 82.3% Fe, 11.2% Ni, 1.0% Co and 0.63% Mg. Since the Mg concentration indicate a relatively high amount (~3.7 wt%) of silicate contamination, we did not use this metal sample for radionuclide analysis.

The non-magnetic fraction (85.5 mg) of the Hamilton sample as well as the bulk Calgary sample (50.25 mg) were dissolved in a mixture of concentrated $HF/HNO_3$ in a Parr Teflon digestion bomb at 125 °C for 24 hr. After dissolution, we separated the Cl fraction (for future $^{36}Cl$ analysis) as AgCl and removed Si from the remaining solution by fuming with $HClO_4$. The residue was dissolved in dilute HCl, the solution was weighed, and a small aliquot (~3%) was taken for chemical analysis by ICP-OES. Results of the chemical analysis for Mg, Al, P, S, K, Ca, Ti, Mn, Fe, Co and Ni are shown in Table 8. The concentrations of most major and minor elements measured in the two samples overlap with those of average bulk compositions of L and LL-chondrites (Wasson and Kallemeyn, 1988). The main siderophile elements (Fe, Co, Ni) of the Hamilton sample are closer to average L, while those of the Calgary sample are closer to average LL composition.

## Short-Lived Radionuclides

**Methods and Data**

We separated Be and Al from the dissolved samples using ion exchange chromatography and acetyl-acetone extraction techniques (e.g., Welten et al., 2001). After separating and purifying the Be fraction, it was precipitated as hydroxide, transferred to quartz vials and ignited in a tube furnace at 850 °C to convert to beryllium oxide (BeO). The BeO powder is mixed with Nb powder and loaded into a stainless-steel sample holder for isotopic analysis by accelerator mass spectrometry (AMS). The $^{10}Be/^{9}Be$ ratios of the two samples were measured by AMS at Purdue

University's Rare Isotope Measurement (PRIME) Lab (Sharma et al., 2000). The measured ratios are corrected for blank levels and normalized to the measured ratios of a $^{10}$Be AMS standard (Nishiizumi et al., 2007). Based on the normalized $^{10}$Be/$^9$Be ratios, the amount of sample dissolved, and the amount of Be carrier added, we calculated the $^{10}$Be concentration in each sample (in atoms/g) and converted these to $^{10}$Be activity in disintegrations per minute per kg (dpm/kg). The $^{10}$Be results are shown in Table 8.

In addition to the $^{10}$Be, cosmogenic radionuclide concentrations have been analyzed by means of non-destructive high purity germanium (HPGe) gamma spectroscopy. The counting efficiencies have been calculated using thoroughly tested Monte Carlo codes. One specimen of Golden (slice from Hamilton fragment) was measured in the underground laboratories at the Laboratori Nazionali del Gran Sasso (LNGS) (Arpesella, 1996; Laubenstein, 2017) for 39.28 days (specimen mass of 13.5305 g, 129 days after the fall). The results for these short-lived radionuclides are given in Table 9.

## Noble Gas Analysis and Cosmic-Ray Exposure (CRE) Age

**Methods**

*Noble gas analysis*

Powdered specimens of Golden were used for noble gas analysis, one of 110 mg from the Hamilton fragment and one of 57 mg from the Calgary fragment from which we extracted two aliquots of ~13-18 mg, referred to as Golden_HAM_A & B and Golden_CAL_A & B, (Table 10), respectively.

We measured the noble gases according to standard measurement routines (Riebe et al., 2017). For preparation, the samples were wrapped in Al foil and preheated at ~110 °C under vacuum for several days to reduce atmospheric contamination. Gas extraction occurred via bulk fusion in one temperature step at ~1700 °C in a Mo crucible for ~30 minutes. We performed one re-extraction step at ~1750 °C for Golden_HAM_A, which demonstrated full gas extraction in the main temperature step. To ascertain the blank level, we measured empty Al foils of the same mass (20.0 mg) as used for the sample wrapping before and after the four aliquot measurements. The blank corrections were <1% for He and Ne, <3.9% for Ar, <27% for Kr and <2.1% for Xe.

*Cosmogenic components*

Cosmogenic nuclides are produced by the irradiation of surface material with cosmic rays and can be used to determine the time between the ejection of a meteoroid from the parent body and its entry into the Earth's shielding atmosphere (e.g. Herzog and Caffee, 2014). To determine the CRE age requires the meteoritic noble gases to be initially resolved into cosmogenic (cos), trapped (tr) and radiogenic (rad) components.

Neon and $^3$He in all samples are purely cosmogenic, while $^4$He exhibits some additional radiogenic contributions (see Figure 18 and Table 10). As we did not detect any trapped isotopic signatures in the Ne (including that of solar wind), the presence of trapped He is safely excluded. The $^{36}$Ar/$^{38}$Ar ratios are slightly higher than the cosmogenic ratio (see Table 11), indicating the presence of trapped Ar. Cosmogenic $^{38}$Ar was thus calculated with a two-component deconvolution between ($^{36}$Ar/$^{38}$Ar)$_{cos}$ = 0.63–0.67 (Wieler, 2002) and ($^{36}$Ar/$^{38}$Ar)$_{tr}$ of 5.32–5.34 (covering the composition of Q and air; Busemann et al., 2000, Nier, 1950).

*Cosmogenic nuclide production rates*

We used the model for ordinary chondrite matrices by Leya and Masarik (2009), hereafter referred to as "LM09") to obtain the production rates of cosmogenic $^3$He, $^{21}$Ne and $^{38}$Ar (Table 12), which are constrained by the ($^{21}$Ne/$^{22}$Ne)$_{cos}$ ratio. Note that systematic uncertainties of 15-20% estimated for the LM09 model itself, as well as uncertainties for elemental concentrations (given in Table 13) are not included in our production rates. For both specimens, the two ($^{21}$Ne/$^{22}$Ne)$_{cos}$ ratios of the respective aliquots are consistent within uncertainty, so we used their error-weighted means, respectively, in the model. The production rates, and hence also ($^{21}$Ne/$^{22}$Ne)$_{cos}$, are a function of the meteoroid's initial size, the depth in which the sample was located (summarized under the term "shielding"), the chemical composition of the meteoroid (represented in the LM09 model as ordinary chondrite "matrix" chemistry) and of the specimen analyzed. For the latter, we used the elemental compositions listed in Table 13.

From the infrasound and ablation modelling, the pre-atmospheric mass is estimated to be ~tens to ~100 kg (see discussion section). We thus constrained the meteoroid's radius to 20–25 cm, allowing for a mass of ~100–200 kg (assuming a spherical shape and a density of 3200 kgm$^{-3}$, see meteorite sample section), i.e., conservatively permitting a deviation by about a factor of two. Note that a meteoroid radius of 20–85 cm for the Calgary specimen and 20–200 cm for the Hamilton specimen would be in agreement with the LM09 model alone. Considering the observational mass constraints, we obtain sample depths of 10–20 cm for the Hamilton sample and 7–10 cm for the Calgary sample.

*Gas retention ages*

The gas retention age determines the closure time of a phase to gas loss (i.e., preventing gas from diffusing through the crystal lattice) and is typically used to trace the thermal history of a meteorite during its residence in the parent body (e.g., Bogard, 2011). We determined this age with the U/Th-He and K-Ar thermochronometers using the detected range of radiogenic $^4$He and $^{40}$Ar for each sample, in combination with K, U and Th concentrations listed in Table 13. We obtained $^4$He$_{rad}$ by assuming that all $^3$He is cosmogenic, ($^3$He/$^4$He)$_{cos}$ =0.163–0.191 (Wieler, 2002) and $^4$He$_{tr}$ is negligible. Radiogenic $^{40}$Ar was derived from the concentration of $^{36}$Ar$_{tr}$ and a ($^{40}$Ar/$^{36}$Ar)$_{tr}$ ratio in the range of 0–295.5 (covering both the compositions of Q which contains essentially no trapped $^{40}$Ar and air; Busemann et al., 2000, Steiger and Jäger, 1977).

## Results and Discussion

*Noble gas composition*

The measured noble gas concentrations and isotopic ratios of Golden are shown in Tables 10, 11, 14 and 15. As described above, the He, Ne and Ar of Golden are predominately cosmogenic, with some contributions from $^4$He$_{rad}$ and $^{40}$Ar$_{rad}$ and a minor proportion of Ar$_{tr}$. The latter is too small to enable us to determine its origin but is typically a mixture of Q and air in ordinary chondrites of petrologic type ≥4 (e.g. Alaerts et al., 1977; 1979). The Kr and Xe compositions are consistent with Q and air and small amounts of cosmogenic Kr and Xe, visible in the isotope ratios $^{78,80}$Kr/$^{84}$Kr and $^{124,126}$Xe/$^{132}$Xe. Figure 19 shows the Xe isotopic compositions of the four aliquots and air normalized to Xe-Q from Busemann et al. (2000). While the aliquots from the Calgary specimen show some contamination with air (AL_A in $^{126,128}$Xe and AL_B in $^{134,136}$Xe), those from the Hamilton specimen seem to be largely unaffected (some air contribution might be present in Hamilton visible in $^{134,136}$Xe, but this could also derive from the decay of $^{244}$Pu and $^{238}$U).

We infer that the difference between samples derives from terrestrial weathering, against which the Hamilton sample was protected by crashing directly into Ruth Hamilton's bedroom. Higher concentrations of $^{84}$Kr and $^{132}$Xe in the Calgary aliquots relative to those of Hamilton (Tables 14 and 15) further support the assumption that the former –found a week after the fall– experienced more weathering than the latter (cf. also Scherer et al., 1994). We also observe an albeit small air contamination in the Kr isotopic compositions of the Calgary aliquots (generally lower $^{80,82,83}$Kr/$^{84}$Kr ratios compared to the Hamilton aliquots, Table 14). We nevertheless conclude that the heavy noble gas inventory is strongly dominated by the Q component, as also the $^{84}$Kr/$^{132}$Xe ratios in the more pristine Hamilton samples are ~0.6/0.7, which is typical for Q-gas (Busemann et al., 2000), while this ratio in air is ~27. The Q component must have survived thermal metamorphism reaching up to ~650–760 °C in L and LL-chondrites of type 5 (Kessel et al. 2007) as was previously observed by Alaerts et al., (1977; 1979). Additionally, all four aliquots show substantial contributions of $^{129}$I-derived $^{129}$Xe excess (Figure 19 and Table 15) as is typical for ordinary chondrites of type 4-6 (e.g. Alaerts et al., 1979; Moniot, 1980). We do not observe any isotopic signatures from solar wind in the light noble gases and hence conclude that Golden is not a regolith breccia, consistent with CT and cut sample observations of Golden lacking brecciation.

*Cosmic ray exposure age*

We calculated the CRE ages $T_x$ for the two specimens separately from the respective error-weighted mean abundances of cosmogenic nuclides ($^3$He, $^{21}$Ne, $^{38}$Ar) and their production rates derived with the LM09 model (Table 12). For the Hamilton sample, the $T_3$ and $T_{21}$ ages are identical, while the $T_{38}$ is slightly higher. However, the latter overlaps with the $T_3$ age of the Calgary sample, which is sightly lower than its consistent $T_{21}$ and $T_{38}$ ages. The generally higher CRE ages of the Calgary specimen match with its shallower sample location derived with the LM09 model. We further measured somewhat lower $(^{21}Ne/^{22}Ne)_{cos}$ ratios for the Calgary aliquots (Fig. 18). This difference could derive from slightly varying plagioclase abundances (on average ~13 vol% in L5 chondrites, Dunn et al., 2010), which is the main host phase of Na in L-chondrites, producing 1.8 times more $^{22}Ne$ than $^{21}Ne$. We find a preferred CRE age for both samples including all three ages (based on mean and standard deviation) of 23±2 Ma for the Hamilton specimen and 27±1 Ma for the Calgary specimen. These ages fall within both the broader peak between 20 and 30 Ma in the CRE age distribution of L-chondrites and the 27-33 Ma peak in LL CRE ages (Herzog and Caffee, 2014).

*Gas retention ages*

The K-Ar ages $T_{40}$ and the U/Th-He ages $T_4$ obtained for the Calgary and the Hamilton specimens individually are shown in Table 16. The $T_4$ ages range from 1.9-2.6 Ga and 2.3-2.7, and the $T_{40}$ ages range from 3.7-3.8 and 3.8-3.9 Ga for the Hamilton and Calgary specimens, respectively.

Both $T_{40}$ ages agree within their respective uncertainties. The $T_4$ ages are considerably younger and Hamilton_A exhibits an outlying minimum $T_4$ age, caused by its smaller concentration of $^4He_{rad}$. The latter could be explained by a varying abundances of apatite and merrillite among the aliquots of the Hamilton specimen, which mainly control the U concentration (e.g., Goreva and Burnett 2001). Assuming a mean particle diameter of 200 µm (estimated based on the examination of the powdered sample under the microscope) and a density of 3200 kgm$^{-3}$ (see above), the 110 mg powder of the Hamilton sample would contain ~8200 particles, of which 0.7 vol% are phosphates (Lewis and Jones 2016), i.e., 57 grains. It thus appears possible that those were unevenly distributed among the two aliquots, which could explain the low $^4$He content in Hamilton_A compared to the other three aliquots (see also mineralogical description above).

The younger U/Th-He ages relative to the K-Ar ages are typically interpreted to represent smaller degassing events late after general accretion and mineral closure, e.g., caused by impacts, which affected the more volatile He more than Ar. These gas retention ages suggest that Golden was not part of the large parent body break up event around 470 Ma ago recorded in many L-chondrites (e.g., Swindle et al., 2014).

*Short-lived Radionuclides*

The measured $^{10}$Be concentration of 19.6 ± 0.2 dpm/kg in the Calgary sample is within the typical range of 15-23 dpm/kg for ordinary chondrites with radii of 10-100 cm (LM09). The measured $^{10}$Be concentration from the dissolved samples using ion exchange chromatography in the non-magnetic fraction of the Hamilton sample is slightly higher at 20.7±0.2 dpm/kg, but this is mainly due to the different chemical composition of the stone fraction vs. the bulk sample. Based on the mass balance of stone (95.5 wt%) and metal (4.5 wt%) and the assumption that the metal fraction contains 5±1 dpm/kg $^{10}$Be (typical for small to medium-sized chondrites), the Hamilton sample has a bulk $^{10}$Be concentration of 20.0 ± 0.2 dpm/kg, within the uncertainty the same as the bulk value of the Calgary sample.

Figure 20 shows that the $^{10}$Be concentrations in the two Golden samples are consistent with an irradiation depth of 5-10 cm in an object of 20-65 cm in radius. The lower end of this size range overlaps with the pre-atmospheric radius of ~20 cm that was derived from the infrasound and fireball modeling of the atmospheric entry. The $^{10}$Be concentration and the shielding sensitive $^{22}$Ne/$^{21}$Ne ratio are also consistent with the expected relationship of $^{10}$Be vs. the $^{22}$Ne/$^{21}$Ne ratio in ordinary chondrites of 20-65 cm in radius, as derived from the model calculations of LM09. Combined, these results indicate that (i) the Golden meteorite had a pre-atmospheric diameter of 40-50 cm with samples coming from a depth of 5-10 cm and (ii) the cosmogenic radionuclides and noble gases were produced under the same shielding conditions, indicating a simple (one-stage) CRE history.

Since the irradiation geometry of the samples did not change during the CRE history of the meteorite, we can use the relationship between the $^{10}$Be and $^{21}$Ne production rates from the model of LM09 combined with the measured $^{10}$Be concentrations to derive the $^{21}$Ne production rate. This approach yields $^{21}$Ne production rates of 0.331 and 0.316 (x10$^{-8}$ cm$^3$ STP/g/Ma) for the Hamilton and Calgary sample, respectively, about 10% lower than the production rates from Table 12. From the $^{10}$Be vs. $^{21}$Ne relationship (which does not use the $^{22}$Ne/$^{21}$Ne ratio for shielding purposes), the $^{21}$Ne-$^{10}$Be CRE age of Golden is ~10% higher than the age inferred from the $^{21}$Ne vs. $^{22}$Ne/$^{21}$Ne relationship.

Among the short-lived radionuclides measured by gamma-ray counting, the activity of $^{60}$Co is rather low (<1.1 dpm/kg). Normalized to the composition of an ordinary H chondrite the measured $^{26}$Al activity is consistent with that expected for a small-sized L/LL chondrite (Bhandari et al., 1989; Bonino et al, 2001; LM09).

When we compare the radionuclide concentrations with cosmic ray production estimations for $^{26}$Al (LM09), $^{60}$Co (Eberhardt et al., 1963; Spergel et al., 1986), $^{54}$Mn (Kohman and Bender, 1967), and $^{22}$Na (Bhandari et al., 1993), the best agreement is obtained (in the sequence of the given isotopes) for radii of 10-120 cm, <20 cm, 13-100 cm and 15-100 cm, all at a depth between 4 cm and 10 cm. The $^{22}$Na/$^{26}$Al ratio of the specimen is (1.87 ± 0.16) in agreement with expectations for such sized meteoroids. Figure 21 shows the expected variation in $^{26}$Al using the Leya et al. (2009) model as a function of depth in L-chondrites of various radii.

The activities of the short-lived radioisotopes, with half-lifes less than the orbital period, represent the production integrated over the last segment of the orbit. The fall of the Golden L/LL5 chondrite occurred during a minimum at the end of solar cycle 24, as indicated by the neutron monitor data (Bartol, 2023). The cosmic ray flux was high in the six months before the fall (Bartol, 2023), so the activities for the very short-lived radionuclides are expected to be rather high (see Table 9). The naturally occurring radionuclides (Table 17) are in good agreement with the average concentrations in ordinary L/LL-chondrites (Wasson and Kallemeyn, 1988).

*Pre-atmospheric mass*

The estimated pre-atmospheric size/mass from all techniques is summarized in Table 18. The noble gas and short-lived nuclides allow a wide range of possible initial sizes, while the infrasound energy and lightcurve and entry modelling are in closer agreement near 17-18 cm radius. It is particularly difficult to reconcile the upper range of radionuclide sizes (meter-range) with the modelling results as the lightcurve would necessarily be much brighter than measured. Note that $^{60}$Co tends to be a robust indicator of small vs. large pre-atmospheric size (Kollar et al., 2006) and our low $^{60}$Co activity suggests a small (<20 cm radius) pre-atmospheric meteoroid.

In particular, our constraint of non-detection from GLM provides a useful upper bound to the meteoroid size. Making the extreme assumption that the entire lightcurve is underestimated by the maximum allowed difference to be consistent with GLM non-detection (assuming a threshold of -15.5 for the ground location of Golden as discussed earlier) the equivalent modelled mass would be ~300 kg or 28 cm radius.

As all techniques overlap in the 17 – 25 cm radius range, our preferred initial mass estimate is in the range of 70 – 200 kg, but masses as high as ~300 kg are nominally allowed within our constraints.

*Orbital Evolution History and Comparison to other L/LL-chondrite Orbits*

In estimating the source location in the main belt of the Golden meteorite, we use the seven escape route (eR) model described by Granvik et al. (2018) which incorporates size-dependent fitting in the calibration per eR. Here we use the smallest H-magnitude bin in their model (H=25) which is appropriate to asteroid diameters of roughly 35 m for an albedo of 0.15. Note that there is a noticeable change in eR probabilities for fixed semi-major axis, eccentricity and inclination (*a,e,i)* combinations as a function of H magnitude and hence the much smaller H magnitude of the Golden progenitor likely has slightly different values.

Using the *a,e,i* as reported in Table 3 (and including the uncertainties) we find that the two major eRs are the $\nu_6$ secular resonance (33±8%) and the Hungaria eR (60±10%). All other eRs have near zero probabilities, except the 3:1 MMR which has a small probability of (7±2%). Note that even adopting larger radiant uncertainties up to 1.5 degrees from nominal values (through inclusion or exclusion of the discordant Sunshine station points) the resulting eR sum of $\nu_6$ and Hungaria remain above 90%.

As the $\nu_6$ and Hungaria are the innermost of the eRs in the Granvik et al. (2018) model, we can say with some confidence that Golden evolved to an Earth-crossing orbit after escaping from the innermost part of the main belt. The high eR from Hungaria reflects the high inclination for Golden, suggesting origin from a parent body with significant inclination. Golden is the only L-chondrite with such a high probability of association with the Hungaria group. While the main Hungaria group are E (or Xe) asteroids (DeMeo and Carry, 2013) unrelated to OCs, Lucas et al. (2017) found that most of the background asteroids in the Hungaria group are S-types and many have spectral affinities to L/LL-chondrites. Our results for Golden support the claim by Lucas et al (2017) that some L-chondrites originate from the inner orbital boundary of the Hungaria family; Golden is the first L/LL-chondrite to show strong orbital indications of just such a parentage.

The Golden meteorite orbit adds more evidence to the emerging picture of the inner main belt as the main source region for L/LL-chondrites, a picture consistent with the origin of NEAs spectrally linked to L and LL-chondrites (e.g., Binzel et al., 2019). Table 19 shows the eR probabilities for all 16 L and LL-chondrites with published orbits to date including Golden, while Table 20 gives their corresponding orbits. The final column in Table 19 is what we term an inner main belt (IMB) probability and is simply the sum of the $\nu_6$ and Hungaria eR probabilities. The most noticeable feature in the table is that with only three exceptions (Dingle Dell, Park Forest and Villabeto) all L and LL (and L/LL) chondrite orbits show an >80% probability of having originated in the inner main belt. Indeed, almost half have IMB eR probabilities >90%. This provides evidence in support of (8) Flora as a possible source for the LL (or maybe L-chondrites) (Vernazza et al., 2008; Dunn et al., 2013) but is in tension with the Gefion family (Nesvorný et al., 2009) as a source, unless long Yarkovsky drift times from the mid-belt location of the Gefion family to the eRs in IMB are invoked. As noted by e.g., Galinier et al. (2023), the HEDs, aubrites and EL-chondrites are also likely linked to asteroid families in the inner main belt.

The notion that the 5:2 or 3:1 MMR, which are closest to the Gefion family, are a major source for L-chondrite delivery is in contradiction with the now considerable number of L-chondrite orbits which show a clear signal associated with escape from the inner main belt. Recent work also suggests many Gefion family members do not have spectral affinity for L-chondrites (McGraw et al., 2018) and that the Gefion family may be older than the 0.5 Ga (Spoto et al., 2015) degassing age found for many L-chondrites, an association which was originally used as a strong argument in favor of Gefion as the LCPB (Nesvorný et al., 2009). This supports earlier studies which also questioned the Gefion as the LCPB link based on a smaller sample of L-chondrite orbits (e.g. Granvik and Brown, 2018; Jenniskens et al., 2019).

Among the L-chondrites with pre-atmospheric orbits, only two are known to have degassing ages consistent with the 470 Ma (Park Forest and Novato) LCPB breakup (Table 20). However, we caution that more than 1/3 of all L-chondrites with orbits have not yet had retention ages reported, so this picture may be incomplete. Park Forest appears to be unusual among L-chondrites orbits as it (and Villalbeto) was likely delivered via the 3:1 MMR (Meier et al., 2017) and both show signs of short gas retention ages (Park Forest near 430-490 Ma, Villalbeto ~700 Ma ago). Whether this common feature points to a different parent body than other L-chondrites originating from the IMB is unclear, a picture further complicated by Novato which comes from the IMB and has a 460-550 Ma degassing age (Jenniskens et al., 2019).

A small subset of ordinary chondrites are classified as transitional L/LL (at least 130 in the Meteoritical Bulletin Database as of January 2023), with a number of other L or LL-chondrites being suspected of being transitional. Two observed falls with calculated orbits appear to be in this transitional group; the 2016 L/LL5 Dingle Dell fall (Devillepoix et al., 2018), and the 1977 L/LL5 Innisfree fall (Halliday et al., 1978). It will be useful to pursue these transitional L/LL meteorites as a potential ordinary chondrite subgroup in further work, as they appear overrepresented amongst the L and LL meteorites with known orbits.

## Conclusions

The Golden (L/LL5) meteorite fall is an unbrecciated, unshocked ordinary chondrite with composition transitional between L and LL. It arrived as a 17-25 cm radius meteoroid with a simple, one stage CRE age of 25±4 Ma. It does not show any evidence for shock resetting associated with the 470 Ma peak found for many L-chondrites, but rather shows a U/Th-He age near 2–2.5 Ga and a K-Ar age of 3.7-3.9 Ga. The pre-atmospheric orbit for Golden indicates its immediate escape region was in the inner main belt, with a high likelihood of association with the Hungaria group. We suggest Golden most likely originates from a parent body in the Hungaria region and is further evidence for multiple immediate parent bodies for L-chondrites as has been suggested by others (e.g., Jenniskens et al., 2019; Devillepoix et al., 2022). In this picture, the contemporary flux of L-chondrites originates from a series of bodies having undergone relatively recent (10s of Ma) collisions in the IMB, consistent with their CRE ages and the Yarkovsky drift time required to reach the $\nu_6$ or Hungaria "ejection" hatches. More broadly, the current suite of L and LL-chondrite orbits indicates that the majority (though not all) of these OCs originate in the IMB.

At present almost 20 L or LL-chondrite orbits are now published or in the process of being reported. Measuring CRE ages and gas retention ages for these unique samples with orbital context (many of which lack such data) is crucial to providing the clues as to their original parents.

**Acknowledgements**

We thank the main mass finder, Ruth Hamilton, for sharing her experience of the Golden fall event, for the loan of her meteorite and for kindly supplying type specimen sample of the fresh meteorite for this research. We thank Hao Qin, Alex Cohen and the Sunshine Village Ski Resort, and Toby Barrett and the Kicking Horse Mountain Resort, for graciously supplying fireball images and related data. We thank numerous witnesses for describing their experiences and landowners Golden Concrete, and April, Doug, and Rob Kinsey for permission to search their properties and logistical support. Funding support for this work was provided by the NASA Meteoroid Environments Office through co-operative agreement 80NSSC21M0073. PGB thanks the Canada Research Chair program, and the Natural Sciences and Engineering Research Council of Canada (NSERC) for funding support. PJAM and RLF acknowledge continuing support from NSERC – Discovery Grants. AN acknowledges the Museum of Ontario Archaeology for facilitating access to the micro-CT scanner. ARH, LTJH, and FC acknowledge support from the Department of Geoscience, University of Calgary. The work conducted at ETH was partially funded by the Swiss National Science Foundation (SNF) with grants 200020_182649 and 51NF40_205606 (NCCR PlanetS). The GFO was funded by the Australian Research Council as part of the Australian Discovery Project scheme (DP170102529, DP200102073), the Linkage Infrastructure, Equipment and Facilities scheme (LE170100106). It receives institutional support from Curtin University, and is supported by software support resources awarded under the Astronomy Data and Computing

Services (ADACS) Merit Allocation Program. We thank Robert Marr and Liane Loiselle for assistance with acquiring EPMA data on the Calgary and Hamilton samples, respectively; Rob Alexander and Jed Baker supported imaging and 3D printing at the University of Calgary. We also thank Jiri Borovicka for a detailed and thoughtful review which helped to improve an earlier version of this manuscript.

## Tables.

**Table 1**. Video and photographic records of the Golden fireball for which calibrations were possible. Station numbering given in the figures is shown in square brackets under the video location where applicable. Videos used for Astrometry [A], Photometry [P] also indicated after location name. For the still images from Vermilion and Mattheis Ranch, the frame rate reflects the shutter frequency.

| Video Location/Name[URL] | Latitude, Longitude (N/W) [degs] | Frame Rate (frames per second) | Duration of fireball signal (sec) | Field of View (HxV) [degs] | Sensor resolution (HxV) [pixels] | Range to endpoint (km) |
|---|---|---|---|---|---|---|
| Lake Louise / Qin [A] | 51.417 , 116.217 | still | | 73x52 | 5584x3728 | 59 |
| Banff, AB /Sunshine [A] | 51.074, 115.761 | 1 | <7.4 | 91x66 | 2048x1536 | 95 |
| Calgary, AB/Saddletowne (Dashcam) [P] | 51.12517, 113.96028 | 30 | 4.8 | | 1280x720 | 217 |
| Calgary, AB/CA000J [A] | 50.910252, 114.039078 | 13 | 3.5 | 152 (all sky) | 1280x720 | 219 |
| Delacour, AB/Neilsen [P] | 51.16498, 113.77128 | 15 | 3.7 | | 1920x1080 | 229 |
| Cranbrook, BC/Cranbrook [A] | 49.51755, 115.74363 | 25 | 3.0 | 178 (all sky) | 640x480 | 235 |
| Mattheis_Ranch, AB/ [A] | 50.8940416667, 111.972911667 | 20 | 0.55 | 152 (All-sky) | 7360x4912 | 375 |
| Vermilion, AB/DFNEXT044 [A,P] | 53.338975, 110.884105 | 20 | 2.75 | 152 (All-sky) | 7360x4912 | 480 |

**Table 2.** Basic atmospheric trajectory information for the Oct 4, 2021, Golden fireball based on calibrated camera solutions. Geographic coordinates are referenced to the WGS84 geoid and are apparent, ground-fixed.

| | Beginning | End |
|---|---|---|
| Height (km) | 83.9 ± 0.1 | 18.5 ± 0.03 |
| Latitude (N) | 51.574° ± 0.001° | 51.310° ± 0.003° |
| Longitude (W) | 117.568° ± 0.002° | 117.048° ± 0.002° |
| Slope | 54.3° ± 0.1° | |
| Azimuth of radiant | 308.5° ± 0.1° | |
| Velocity (km/s) | 17.93 ± 0.08 | < 4 |
| Trail Length/Duration | 81.4 km /6.4 s | |
| Time (UT) | 05h33m44.02s | 05h33m50.4s |

**Table 3.** Heliocentric orbit for the fireball producing the Golden meteorite. All angular coordinates are referenced to J2000.0. $V_\infty$ refers to the speed of the fireball relative to the Earth's surface prior to significant (>10 m/s velocity difference) atmospheric deceleration, which for Golden occurred at a height of 70 km. Note the Tisserand range represents the 95% confidence interval.

| | |
|---|---|
| $\alpha_r$ | 274.75 ± 0.14° |
| $\delta_r$ | 59.52 ± 0.18° |
| $V_\infty$ | 17.9 ± 0.08 km/s |
| $V_G$ | 14.1 ± 0.1 km/s |
| $\alpha_G$ | 265.86 ± 0.14 ° |
| $\delta_G$ | 57.16 ± 0.2° |
| a | 1.58 ± 0.02 A.U. |
| e | 0.366 ± 0.006 |
| i | 23.5 ± 0.13° |
| $\omega$ | 177.04 ± 0.13° |
| $\Omega$ | 190.9243° |
| q | 1.00003 ± 0.00002 A.U. |
| Q | 2.16 ± 0.03 A.U. |
| $T_j$ | 4.20 <$T_j$< 4.27 |

**Table 4.** Covariance matrix for Golden meteorite orbital elements.

|   | e | q (AU) | Tp (JD) | Ω (degs) | ω (degs) | inc (degs) |
|---|---|---|---|---|---|---|
| e | 1.74091E-05 | -5.41154E-08 | -0.000271509 | 6.32452E-10 | -0.000324545 | -7.68603E-06 |
| q (AU) | -5.41154E-08 | 4.05847E-10 | 2.1931E-06 | -1.16957E-11 | 2.56171E-06 | 8.86193E-08 |
| Tp (JD) | -0.000271509 | 2.1931E-06 | 0.01194186 | -6.53187E-08 | 0.01393162 | 0.000487951 |
| Ω (degs) | 6.32452E-10 | -1.16957E-11 | -6.53187E-08 | 1.31394E-11 | -7.51544E-08 | -1.04133E-07 |
| ω (degs) | -0.000324545 | 2.56171E-06 | 0.01393162 | -7.51544E-08 | 0.01625838 | 0.000563305 |
| inc (degs) | -7.68603E-06 | 8.86193E-08 | 0.000487951 | -1.04133E-07 | 0.000563305 | 0.000838381 |

**Table 5.** Representative fragmentation history for the Golden fireball which reproduces the observed lightcurve and lag.

| Frag Type | Height (km) | Time (sec) | Mass (%) | Eros Coeff $s^2km^{-2}$ | Dyn Pres (MPa) | Min Grain mass (kg) | Max Grain Mass (kg) | Velocity (km/s) | Parent Mass (kg) | Fragment Mass (kg) |
|---|---|---|---|---|---|---|---|---|---|---|
| START | 180 | -9.05 |   | 0 |   |   |   |   |   |   |
| EF | 65 | 1.05 | 5 | 0.2 | 0.04 | $10^{-6}$ | $10^{-5}$ | 17.9 | 69.9 | 3.5 |
| EF | 34.5 | 3.25 | 10 | 0.2 | 2.3 | $10^{-6}$ | $10^{-3}$ | 16.1 | 62.3 | 6.2 |
| EF | 31.1 | 3.53 | 85 | 0.1 | 3.3 | $10^{-6}$ | $10^{-1}$ | 14.9 | 54.1 | 46.0 |
| E F | 24.0 | 4.34 | 80 | 0.06 | 2.5 | $10^{-3}$ | $10^{-1}$ | 7.5 | 6.9 | 0.99 |
| F | 19.4 | 6.55 |   |   |   |   |   | <3 |   | 1.3 |

Note: Here the fragmentation type is either F, where a single fragment is produced which continues to ablate as a single body or EF where a fragment is produced, continues to ablate as a single body and loses mass mainly through grain erosion where the ablation from the fragment is determined by the erosion coefficient and each grain ablates following single body theory (see Borovička et al. (2020) for more details) contributing to the light production. The height of fragmentation is shown as is the dynamic pressure at the fragmentation point and the zero time is the same as shown in Figure 11. For EF the min and max grain masses are shown. The main body (parent) meteoroid mass at each fragmentation epoch is given as is the released fragment mass. In this scenario a total of two main fragments of 1.3 and 0.9 kg of material survive to reach the ground (last two rows), in addition to numerous smaller fragments from higher release heights. Here the initial mass is 70 kg. The mass (%) column indicates what fraction of the parent mass remaining at a particular height is partitioned into fragmentation, with the total fragmentation mass released shown in the final column.

**Table 6**. Summary of infrasound signal characteristics from the Golden bolide.

| Station | Range (km) | Back Az (Th) | Back Az (Obs) | Celerity | Max Amp(Pa) | SD. Max. Amp (Pa) | Max P-P Amp (Pa) | SD P-P Amp (Pa) | Period at maximum amplitude (s) | SD Period (s) | Period from PSD Max | Dur (s) | Total signal energy (Pa2) | Integr. Signal-2-noise ratio | SCI (m/s) |
|---|---|---|---|---|---|---|---|---|---|---|---|---|---|---|---|
| I10CA | 1483 | 282.8 | 278.6 | 0.306 | 0.01 | 0.005 | 0.02 | 0.01 | 1.09 | 0.05 | 1.02 | 189 | 0.0014 | 2.3 | 25.6 |
| I56US | 337 | 0.3 | 1.8 | 0.296 | 0.043 | 0.007 | 0.064 | 0.014 | 0.6 | 0.03 | 0.60 | 195 | 0.0015 | 15.5 | 7.9 |

Note: Included for each of the two station arrays are the theoretical and observed backazimuth (measured positive from North) of the wave arrival, the average signal speed (celerity) and the maximum and peak-to-peak pressure amplitude (as well as their respective standard deviations (SD) across the array) as well as the period at maximum amplitude from zero crossings. Also shown is the dominant period based on the peak in the power spectral density and the total bolide acoustic energy and integrated signal to noise of the bolide signal. The SCI is the stratospheric circulation index and represents the mean wind between 30-50 km altitude from the source to the receiver with positive values indicating that the wind vector is pointing along the propagation direction (i.e. with wind returns). Note that the signal was bandpassed from 0.7 – 2.5 Hz for I10CA and from 0.5-8 Hz for I56US to maximize the signal to noise ratio.

**Table 7.** Summary of period measurements and uncertainties from the Golden fireball.

| Station | Zero crossing Period | PSD max period | PMCC Max period | PMCC Pseudo-period | Average | Std Dev |
|---|---|---|---|---|---|---|
| I10CA | 1.09 | 1.02 | 1.10 | 0.90 | 1.03 | 0.09 |
| I56US | 0.60 | 0.60 | 0.75 | 0.70 | 0.66 | 0.08 |

**Table 8**. Measured concentrations of major elements (in wt%), minor elements (in ppm) and cosmogenic $^{10}$Be (dpm/kg) in Golden L/LL5 chondrite.

| Element/ nuclide | unit | Golden-HA (85.53 mg) | Golden-HA bulk | Golden-CA (50.25 mg) | L* | LL* |
|---|---|---|---|---|---|---|
| Mg | wt% | 15.6 | 14.9 | 15.7 | 14.9 | 15.3 |
| Al | wt% | 1.14 | 1.09 | 1.10 | 1.22 | 1.19 |
| P | ppm | 1100 | 1080 | 1090 | 950 | 850 |
| S | wt% | 3.0 | 2.9 | 2.1 | 2.2 | 2.3 |
| K | ppm | 1010 | 965 | 900 | 825 | 790 |
| Ca | wt% | 1.26 | 1.20 | 1.26 | 1.31 | 1.30 |
| Ti | ppm | 850 | 810 | 530 | 630 | 620 |
| Mn | wt% | 0.265 | 0.254 | 0.261 | 0.257 | 0.262 |
| Fe | wt% | 18.7 | 21.5 | 19.6 | 21.5 | 18.5 |
| Co | ppm | 120 | 570 | 430 | 590 | 490 |
| Ni | wt% | 0.71 | 1.18 | 1.03 | 1.20 | 1.02 |
| $^{10}$Be | (dpm/kg) | 20.7±0.2 | 20.0±0.2 | 19.6±0.2 | - | - |

HA = Hamilton fragment (Univ. Western Ontario); bulk values are based on non-magnetic fraction of 85.5 mg and magnetic (metal) fraction of 4.0 mg (87% Fe, 11% Ni, 1.0% Co, 0.6% Mg)

CA = bulk sample, Calgary fragment (Univ. of Calgary)

*Average bulk composition for L and LL-chondrites from Wasson and Kallemeyn (1988)

**Table 9.** Massic activities (corrected to date of fall of the meteorite October 4$^{th}$, 2021) of cosmogenic radionuclides (in dpm kg$^{-1}$) in the specimens of the Golden stone measured by non-destructive gamma-ray spectroscopy. Errors include a 1σ uncertainty of 10% in the detector efficiency calibration.

| Nuclide | Half-life | Golden (13.5305 g) |
|---|---|---|
| $^{7}$Be | 53.22 d | 70 ± 20 |
| $^{58}$Co | 70.83 d | 5 ± 1 |
| $^{56}$Co | 77.236 d | 10 ± 2 |
| $^{46}$Sc | 83.787 d | 10 ± 2 |
| $^{57}$Co | 271.8 d | 9 ± 1 |
| $^{54}$Mn | 312.3 d | 79.6 ± 6.2 |
| $^{22}$Na | 2.60 y | 98.2 ± 5.8 |
| $^{60}$Co | 5.27 y | < 1.1 |
| $^{44}$Ti | 60 y | 2 ± 1 |
| $^{26}$Al | 7.17x10$^{5}$ y | 53.6 ± 3.3 |

**Table 10.** Masses used for the noble gas measurements (in mg), as well as He and Ne concentrations (in 10$^{-8}$ cm$^{3}$/g STP) and isotopic ratios.

| | Mass | $^{4}$He | $^{3}$He/$^{4}$He x10$^{4}$ | $^{20}$Ne | $^{20}$Ne/$^{22}$Ne | $^{21}$Ne/$^{22}$Ne |
|---|---|---|---|---|---|---|
| **Golden_HAM A** | 18.748±0.018 | 678±18 | 433.0±2.8 | 7.046±0.043 | 0.8241±0.0044 | 0.9042±0.0037 |
| **Golden_HAM B** | 16.091±0.015 | 861±19 | 358.1±2.8 | 7.042±0.048 | 0.8295±0.0050 | 0.9127±0.0038 |
| **Golden_CAL A** | 13.032±0.019 | 893±22 | 401.4±3.3 | 8.819±0.054 | 0.8251±0.0044 | 0.8985±0.0036 |
| **Golden_CAL B** | 18.475±0.025 | 861±22 | 410.4±3.3 | 8.605±0.051 | 0.8303±0.0041 | 0.8995±0.0039 |
| **cosmogenic** | - | - | 1629–1911[a] | - | 0.704–0.933[a] | 0.800–0.952[a] |

*[a] Wieler 2002*

**Table 11.** Ar concentrations (in $10^{-8}$ cm$^3$/g STP) and isotopic ratios for both samples.

|  | $^{36}$Ar | $^{36}$Ar/$^{38}$Ar | $^{40}$Ar/$^{36}$Ar |
|---|---|---|---|
| **Golden_HAM A** | 1.883±0.010 | 1.4255±0.0085 | 2482±23 |
| **Golden_HAM B** | 1.984±0.011 | 1.5333±0.0093 | 2511±26 |
| **Golden_CAL A** | 1.964±0.015 | 1.483±0.012 | 2501±33 |
| **Golden_CAL B** | 1.8450±0.0077 | 1.3847±0.0073 | 2591±26 |
| **Cosmogenic** | - | 0.63–0.67[a] | - |

**Table 12.** Cosmogenic isotope concentrations (in $10^{-8}$ cm$^3$/g STP), deduced shielding depths (in cm), production rates $P_x$ (in $10^{-8}$ cm$^3$/(g*Ma)) and cosmic ray exposure ages $T_x$ (in Ma).

|  | $^3$He$_{cos}$ | $^{21}$Ne$_{cos}$ | $^{38}$Ar$_{cos}$ | shielding depth |
|---|---|---|---|---|
| **Golden_HAM A** | 38.88±0.15 | 7.73±0.04 | 1.10±0.13 | 10–20 |
| **Golden_HAM B** | 38.68±0.18 | 7.75±0.04 | 1.047±0.011 | |
| **Golden_CAL A** | 46.36±0.24 | 9.60±0.05 | 1.086±0.016 | 7–10 |
| **Golden_CAL B** | 45.99±0.20 | 9.32±0.05 | 1.1208±0.0090 | |

*(continued)*

|  | $P_3$ | $P_{21}$ | $P_{38}$ | $T_3$ | $T_{21}$ | $T_{38}$ | Preferred |
|---|---|---|---|---|---|---|---|
| **HAM A** | 1.824±0.043 | 0.362±0.016 | 0.042±0.001 | 21±1 | 21±1 | 25±1 | 23±2 |
| **HAM B** | | | | | | | |
| **CAL A** | 1.780±0.010 | 0.345±0.006 | 0.0404±0.0004 | 25.9±0.2 | 27.5±0.6 | 27.3±0.4 | 27±1 |
| **CAL B** | | | | | | | |

**Table 13.** Elemental compositions for Golden used to calculate the cosmogenic nuclide production rates, as well as gas retention ages. Values without further indication were measured and described in Table 8. Values designated with * were taken from Wasson and Kallemeyn (1988) and averaged from L and LL chondrites. The uncertainties of major elements are estimated to 2–3% and for minor elements 5–10%.

| Element | unit | Golden_HAM | Golden_CAL |
|---|---|---|---|
| O  | wt% | 37*    | 37*    |
| Na | wt% | 0.7*   | 0.7*   |
| Mg | wt% | 14.9   | 15.7   |
| Al | wt% | 1.09   | 1.1    |
| Si | wt% | 18.7*  | 18.7*  |
| S  | wt% | 2.9    | 2.1    |
| P  | ppm | 1080   | 1090   |
| Ca | wt% | 1.2    | 1.26   |
| K  | ppm | 965    | 900    |
| Ti | ppm | 810    | 530    |
| Mn | wt% | 0.25   | 0.26   |
| Fe | wt% | 21.5   | 19.6   |
| Ni | wt% | 1.18   | 1.03   |
| Co | ppm | 570    | 430    |
| U  | ppm | 0.013* | 0.013* |
| Th | ppm | 0.043* | 0.043* |

**Table 14.** Kr concentrations (in $10^{-10}$ cm$^3$/g STP) and isotopic ratios ($^{84}$Kr = 100).

|  | $^{84}$Kr | $^{78}$Kr/$^{84}$Kr | $^{80}$Kr/$^{84}$Kr | $^{82}$Kr/$^{84}$Kr | $^{83}$Kr/$^{84}$Kr | $^{86}$Kr/$^{84}$Kr |
|---|---|---|---|---|---|---|
| Golden_HAM A | 0.907±0.018 | 0.86±0.15 | 8.48±0.50 | 22.74±0.99 | 23.88±0.89 | 32.3±2.0 |
| Golden_HAM B | 0.909±0.030 | 0.73±0.25 | 8.83±0.56 | 23.3±1.4 | 23.1±1.2 | 31.9±2.3 |
| Golden_CAL A | 2.678±0.031 | 0.802±0.072 | 4.98±0.19 | 21.01±0.53 | 20.88±0.39 | 31.24±0.94 |
| Golden_CAL B | 2.645±0.032 | 0.762±0.061 | 5.28±0.25 | 21.29±0.44 | 21.56±0.48 | 31.15±0.73 |

**Table 15.** Xe concentrations (in $10^{-10}$ cm$^3$/g STP) and isotopic ratios ($^{132}$Xe = 100).

|  | $^{132}$Xe | $^{124}$Xe/$^{132}$Xe | $^{126}$Xe/$^{132}$Xe | $^{128}$Xe/$^{132}$Xe | $^{129}$Xe/$^{132}$Xe |
|---|---|---|---|---|---|
| Golden_HAM A | 1.360±0.016 | 0.539±0.042 | 0.572±0.021 | 8.26±0.18 | 130.6±1.0 |
| Golden_HAM B | 1.416±0.019 | 0.499±0.036 | 0.559±0.033 | 8.17±0.20 | 134.9±1.5 |
| Golden_CAL A | 1.959±0.022 | 0.450±0.022 | 0.374±0.030 | 7.87±0.12 | 112.7±1.0 |
| Golden_CAL B | 1.888±0.021 | 0.467±0.048 | 0.483±0.023 | 7.78±0.21 | 115.10±0.92 |

*(continued)*

|  | $^{130}$Xe/$^{132}$Xe | $^{131}$Xe/$^{132}$Xe | $^{134}$Xe/$^{132}$Xe | $^{136}$Xe/$^{132}$Xe |
|---|---|---|---|---|
| HAM A | 16.50±0.19 | 80.7±1.3 | 38.56±0.41 | 32.71±0.45 |
| HAM B | 15.93±0.21 | 81.8±1.0 | 38.06±0.52 | 31.62±0.47 |
| CAL A | 15.78±0.17 | 80.95±0.81 | 38.31±0.50 | 31.65±0.46 |
| CAL B | 15.97±0.23 | 81.10±0.98 | 39.10±0.26 | 32.88±0.56 |

**Table 16.** Radiogenic isotope concentrations (in $10^{-8}$ cm$^3$/g STP), as well gas retention ages (in Ga)

|  | $^{4}$He$_{rad}$ | $^{40}$Ar$_{rad}$ | T$_4$ | T$_{40}$ |
|---|---|---|---|---|
| Golden_HAM A | 678±18 | 4500±150 | 1.9–2.6 | 3.7–3.8 |
| Golden_HAM B | 861±19 | 4790±140 | | |
| Golden_CAL A | 893±22 | 4720±150 | 2.3–2.7 | 3.8–3.9 |
| Golden_CAL B | 861±22 | 4610±160 | | |

**Table 17**. Concentration of primordial radionuclides (ng g$^{-1}$ for U and Th chains and dpm kg$^{-1}$ for $^{40}$K) in the specimens of the Golden stone measured by non-destructive gamma-ray spectroscopy. Errors include a 1σ uncertainty of 10% in the detector efficiency calibration.

| Nuclide | Golden (13.5305 g) |
|---|---|
| $^{238}$U | 11.9 ± 1.0 |
| $^{232}$Th | 43.5 ± 3.5 |
| $^{40}$K | 870 ± 90 |

**Table 18.** Summary of pre-atmospheric mass (sizes) for the Golden meteorite from different methods. Here radius – mass assumes a spherical body and mean bulk density of 3200 kgm$^{-3}$

| Technique | Mass | Radius (cm) |
|---|---|---|
| Infrasound | $78^{+157}_{-65}$ | 18±8 |
| Lightcurve/Modelling | ~70 | ~17 |
| $^3$He, $^{21}$Ne and $^{38}$Ar | 100-10$^5$ | 20-200 |
| $^{10}$Be | 100-4×10$^3$ | 20-65 |
| $^{26}$Al | 13 – 2.4×10$^4$ | 10-120 |
| $^{60}$Co | <100 | <20 |
| $^{54}$Mn | 30 – 1.3×10$^4$ | 13-100 |
| $^{22}$Na | 50 - 1.3×10$^4$ | 15-100 |

**Table 19.** Escape Region (ER) Probabilities for Golden and other L/LL chondrites using the Granvik et al. (2018) model. References for orbit and CRE ages: (1) This study; (2) Devillepoix et al. (2018); (3) Anderson S. L. et al. (2023); (4) Halliday et al. (1981); (5) Goswami et al. (1978); (6) Borovička et al. (2013); (7) Righter et al. (2015); (8) Brown et al. (2004); (9) Meier et al. (2017); (10) Jenniskens et al. (2019); (11) Gardiol et al. (2020); (12) Andrade et al. (2022); (13) Shrbený et al. (2022); (14) Spurný et al. (2010); (15) Bischoff et al. (2011); (16) Jenniskens et al. (2014); (17) Spurný et a.l (2020); (18) Spurný et al. (2016); (19) Bischoff et al. (2017); (20) Jenniskens et al. (2020); (21) Trigo-Rodriguez et al (2006) (22) Devillepoix et al. (2022). Highlighted falls have gas retention ages consistent with the 470 Ma shock event associated with the LCPB.

| ER→ Meteorite | CRE Ma | $\nu_6$ | $\sigma$ | 5:2 | 2:1 | Hungaria | $\sigma$ | 3:1 | $\sigma$ | Phocaeas | $\sigma$ | IMB Sum |
|---|---|---|---|---|---|---|---|---|---|---|---|---|
| **Golden (L/LL5)[1]** | **25[1]** | **0.33** | **0.08** | **0** | **0** | **0.60** | **0.10** | **0.07** | **0.02** | **0.00** | **0.02** | **0.93** |
| Dingle Dell (L/LL6)[2] | 9[3] | 0.65 | 0.06 | 0 | 0 | 0.03 | 0.01 | 0.30 | 0.06 | 0.00 | 0.00 | 0.68 |
| Innisfree (L5)[4] | 27[5] | 0.59 | 0.08 | 0 | 0 | 0.27 | 0.07 | 0.14 | 0.03 | 0.00 | 0.01 | 0.86 |
| Chelyabinsk (LL5)[6] | 1.2[7] | 0.80 | 0.05 | 0 | 0 | 0.07 | 0.03 | 0.13 | 0.03 | 0.00 | 0.00 | 0.87 |
| Park Forest (L5)[8] | 14[9] | 0.20 | 0.04 | 0 | 0 | 0.01 | 0.00 | 0.52 | 0.08 | 0.00 | 0.00 | 0.21 |
| Creston (L5/L6)[10] | 45[10] | 0.72 | 0.06 | 0 | 0 | 0.15 | 0.05 | 0.14 | 0.03 | 0.00 | 0.00 | 0.87 |
| Cavezzo (L5-Anom)[11] | | 0.81 | 0.05 | 0 | 0 | 0.12 | 0.04 | 0.07 | 0.02 | 0.00 | 0.00 | 0.93 |
| Traspena (L5)[12] | | 0.75 | 0.06 | 0 | 0 | 0.17 | 0.06 | 0.08 | 0.02 | 0.00 | 0.01 | 0.92 |
| Antonin (L5)[13] | | 0.63 | 0.07 | 0 | 0 | 0.25 | 0.07 | 0.12 | 0.03 | 0.00 | 0.01 | 0.88 |
| Jesenice (L6)[14] | 4[15] | 0.71 | 0.07 | 0 | 0 | 0.22 | 0.07 | 0.07 | 0.02 | 0 | 0 | 0.93 |
| Novato (L6)[16] | 9[16] | 0.85 | 0.04 | 0.0 | 0.00 | 0.07 | 0.03 | 0.08 | 0.02 | 0.00 | 0.00 | 0.92 |
| Zdar nad Sazavou (L3.9)[17] | | 0.75 | 0.05 | 0 | 0 | 0.05 | 0.02 | 0.20 | 0.04 | 0 | 0 | 0.80 |
| Stubenberg (LL6)[18] | 36[19] | 0.71 | 0.07 | 0 | 0 | 0.20 | 0.06 | 0.09 | 0.02 | 0 | 0 | 0.91 |
| Dishchii'bikoh (LL7)[20] | 11[20] | 0.62 | 0.07 | 0 | 0 | 0.22 | 0.06 | 0.16 | 0.03 | 0 | 0 | 0.84 |
| Villalbeto(L6)[21] | 48[21] | 0.42 | 0.06 | 0.0 | 0.0 | 0.02 | 0.01 | 0.50 | 0.07 | 0.00 | 0.00 | 0.45 |
| Madura Cave (L5)[22] | | 0.76 | 0.07 | 0.0 | 0.00 | 0.20 | 0.06 | 0.04 | 0.01 | 0.00 | 0.00 | 0.96 |

**Table 20.** Meteorites with known orbits and L/LL, L or LL classification. References: (1) Halliday et al. (1981); (2) Borovička et al. (2013); (3) Spurný et al. (2016); (4) Jenniskens et al. (2020); (5) Devillepoix et al. (2018); (6) This work; (7) Anderson et al. (2023); (8) Bischoff et al. (2017); (9) Righter et al. (2015); (10) Brown et al. (2004); (11) Meier et al. (2017); (12) Jenniskens et al. (2015); (13) Gardiol et a.l (2020); (14) Andrade et al. (2022); (15) Shrbený et al. (2022); (16) Spurný et al. (2010); (17) Bischoff et al. (2011); (18) Jenniskens et al. (2014); (19) Spurný et al. (2020); (20) Llorca et al. (2005); (21) Trigo-Rodriguez et al (2006); (22) Devillepoix et al. (2022). Highlighted falls have gas retention ages consistent with the 470 Ma shock event associated with the LCPB.

| Meteorite Name | Date of Fall (UT) | Meteorite Type | Recovered Mass (kg) | $V_\infty$ (km/s) | $M_{init}$ kg | a (AU) | e | inc (J2000) | q AU | CRE age (Ma) | Ref |
|---|---|---|---|---|---|---|---|---|---|---|---|
| **Golden** | **2021-10-04** | **L/LL5** | **2.2** | **17.9** | **25-75** | **1.58** | **0.366** | **23.5** | **1.00172** | **25±2** | **6** |
| Innisfree | 1977-02-06 | L5 | 4.58 | 14.5 | 30 | 1.87 | 0.47 | 12.2 | 0.9911 | 27 | 1 |
| Chelyabinsk | 2013-02-15 | LL5 | >100 | 19.03 | 1.20E+07 | 1.72 | 0.571 | 4.98 | 0.73788 | 1.2±1 | 2,9 |
| Stubenberg | 2016-03-06 | LL6 | 1.473 | 13.91 | 450 | 1.525 | 0.395 | 2.07 | 0.92263 | 36±3 | 3,8 |
| Dishchii'bikoh | 2016-06-02 | LL7 | 0.079 | 16.56 | 1050 | 1.129 | 0.205 | 21.24 | 0.89756 | 11±3 | 4 |
| Dingle Dell | 2016-10-31 | L/LL5 | 1.15 | 15.443 |  | 2.254 | 0.5904 | 4.051 | 0.92324 | 9.3 | 5,7 |
| Park Forest | 2003-03-27 | (L5) | 18 | 19.5 | 7000 | 2.53 | 0.68 | 3.2 | 0.8096 | 14 | 10,11 |
| Creston | 2016-10-31 | L5/L6 | 0.68 | 16 | 50 | 1.3 | 0.41 | 4.2 | 0.767 | 45 | 12 |
| Cavezzo | 2020-01-01 | L5-Anom | 1.5 | 12.8 | 3.50 | 1.82 | 0.46 | 4 | 0.9828 |  | 13 |
| Traspena | 2021-01-18 | L5 | 0.527 | 16.42 | 2620 | 1.125 | 0.386 | 4.55 | 0.69075 |  | 14 |
| Antonin | 2021-07-15 | L5 | 0.352 | 17.68 | 200 | 1.1269 | 0.2285 | 24.22 | 0.8694 |  | 15 |
| Jesenice | 2009-04-09 | L6 | 3.6 | 13.8 | 170 | 1.75 | 0.43 | 9.6 | 0.9975 | 4 | 16,17 |
| Novato | 2012-10-18 | L6 | 0.36 | 13.67 | 80 | 2.09 | 0.526 | 5.5 | 0.99066 | 9 | 18 |
| Zdar nad Sazavou | 2014-12-09 | L3.9 | 0.087 | 21.89 | 150 | 2.093 | 0.6792 | 2.796 | 0.6714344 |  | 19 |
| Villalbeto | 2004-01-04 | L6 | 3.5 | 16.9 | 600 | 2.3 | 0.63 | 0 | 0.851 | 20 | 20,21 |
| Madura Cave | 2020-06-19 | L5 | 1.072 | 14 | 64 | 0.89 | 0.33 | 0.12 | 0.6 |  | 22 |

# Figures

**Figure 1.** Regional weather overview showing the ground projection of the Golden fireball trajectory (red line – arrow indicating the direction of motion). This image was captured by GOES-17 on Oct 4, 2021, at 052031 UTC in the 1.4μm wavelength passband.

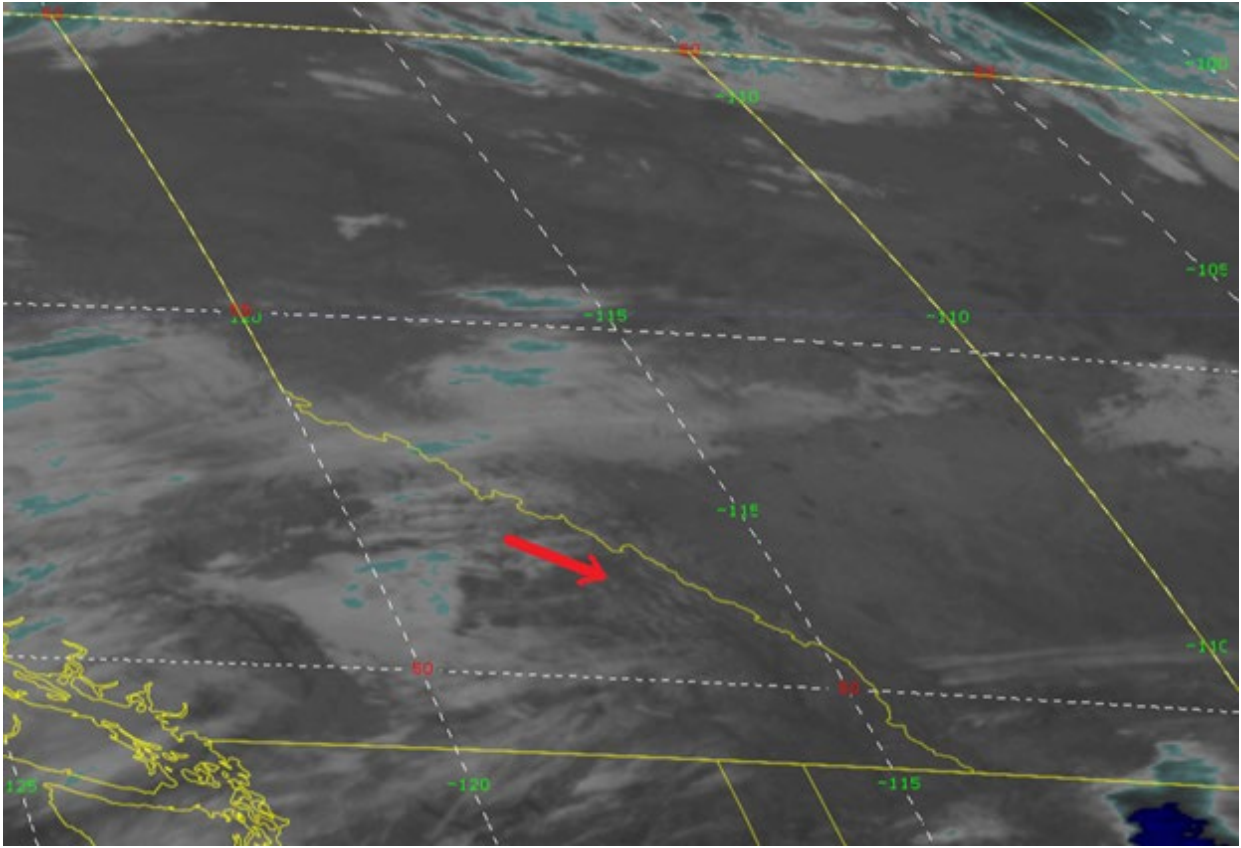

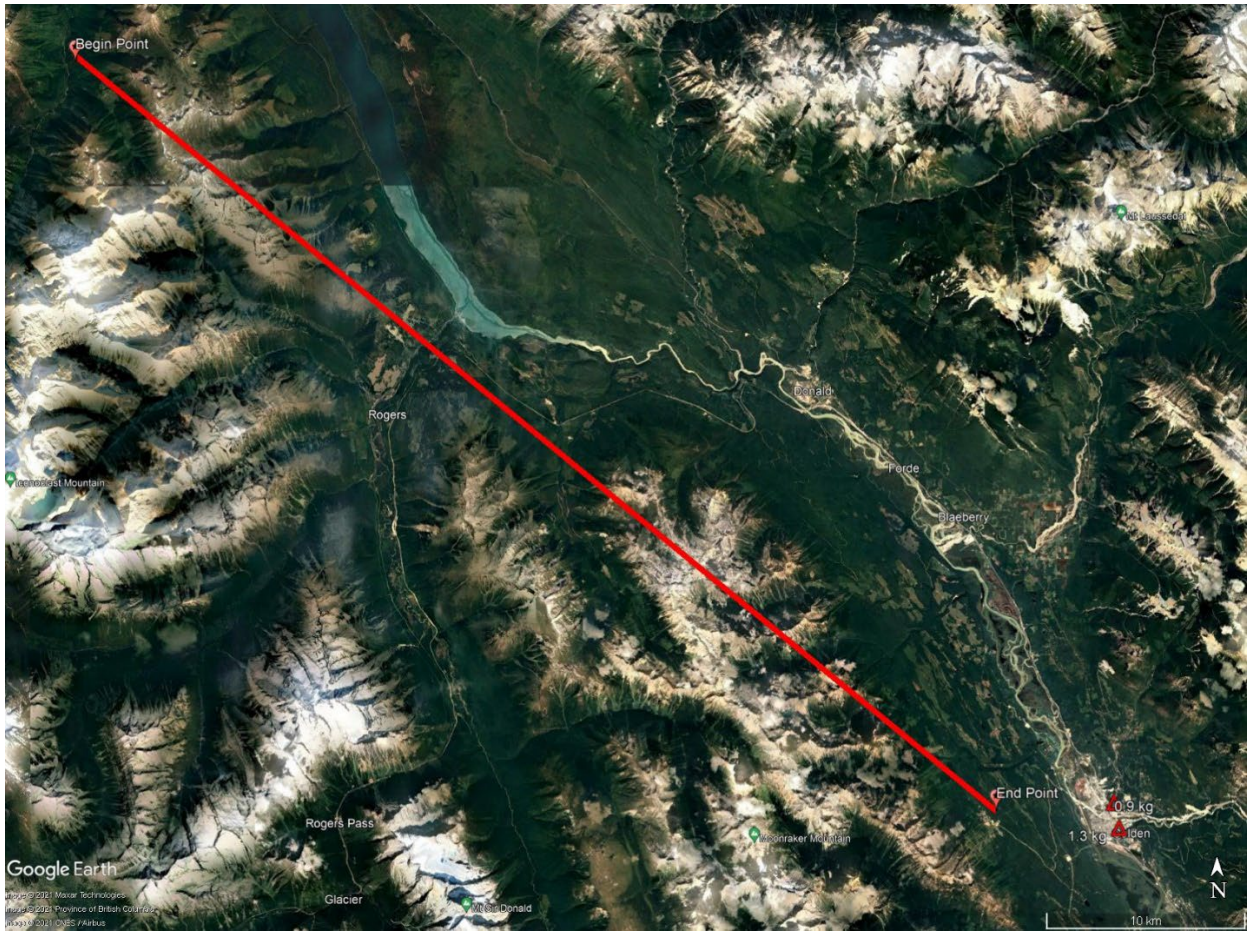

**Figure 2.** Local map overview showing the ground projection of the Golden fireball trajectory from NW to SE (red line) over mountainous terrain west of the Columbia River, BC, Canada together with the location of both recovered meteorites (red triangles) at the Golden town site, bottom right.

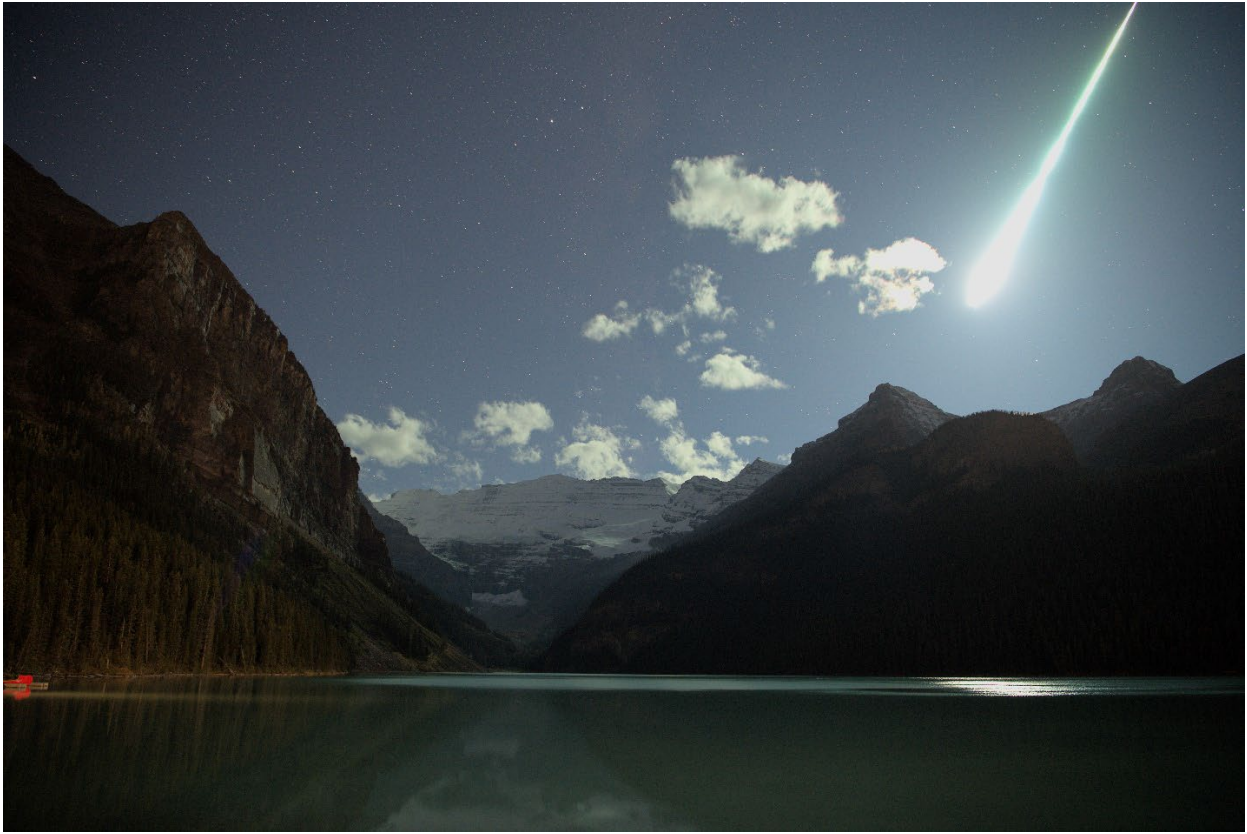

**Figure 3**. Time exposure capturing the early and mid-portion of the Golden fireball along with reference starfield, looking SW from Lake Louise (image courtesy Hao Qin).

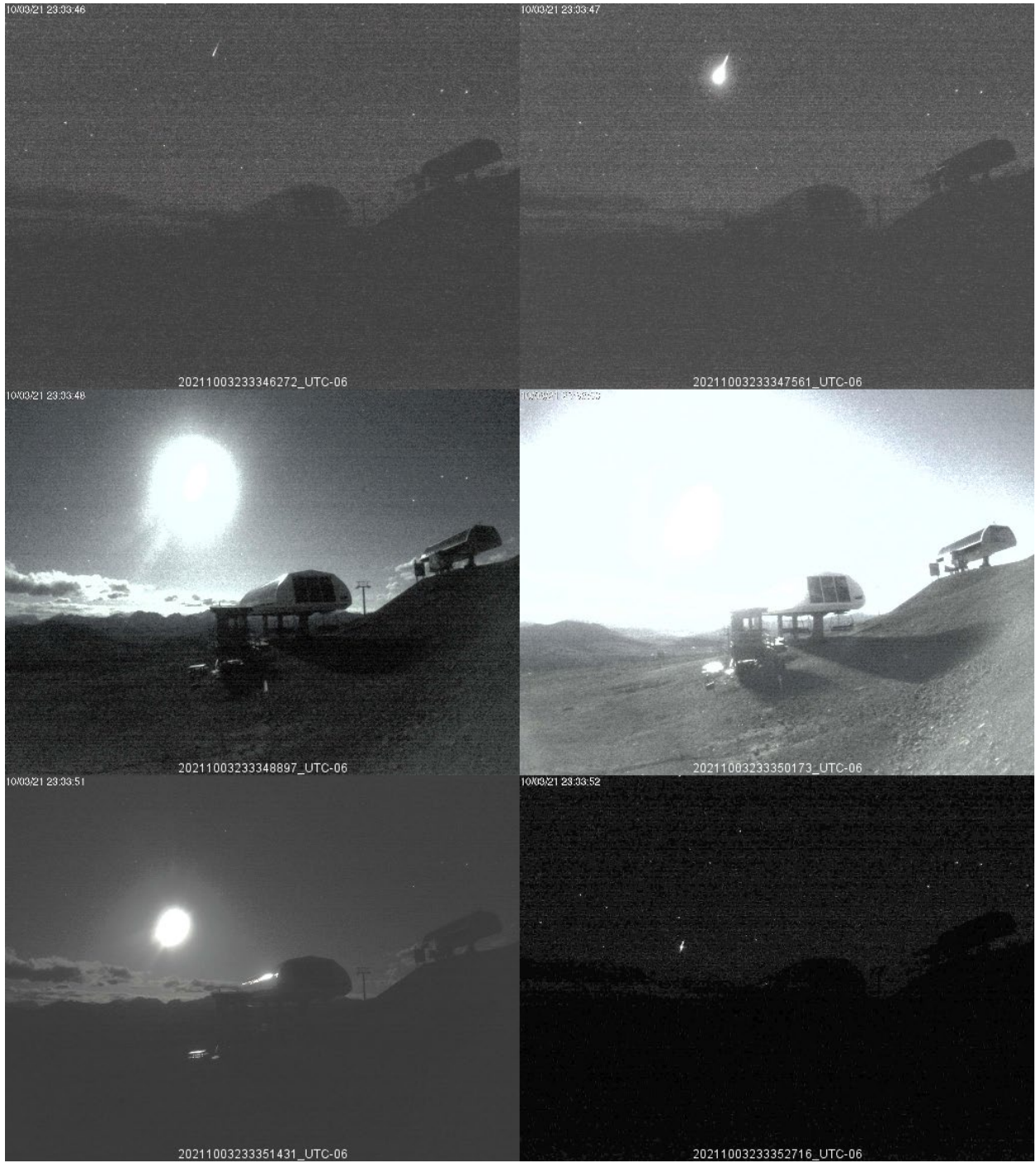

**Figure 4.** Still frames from the IP video recorded at the Banff Ski Resort – Sunshine mountain camera. The camera look direction (towards the NW) was only 25 degrees from the fireball flight direction. Images are recorded at intervals from 1.25 – 1.33 sec with timestamps given at the bottom of the image to one ms precision. Note the appearance of two distinct fragments in the final image at the lower right.

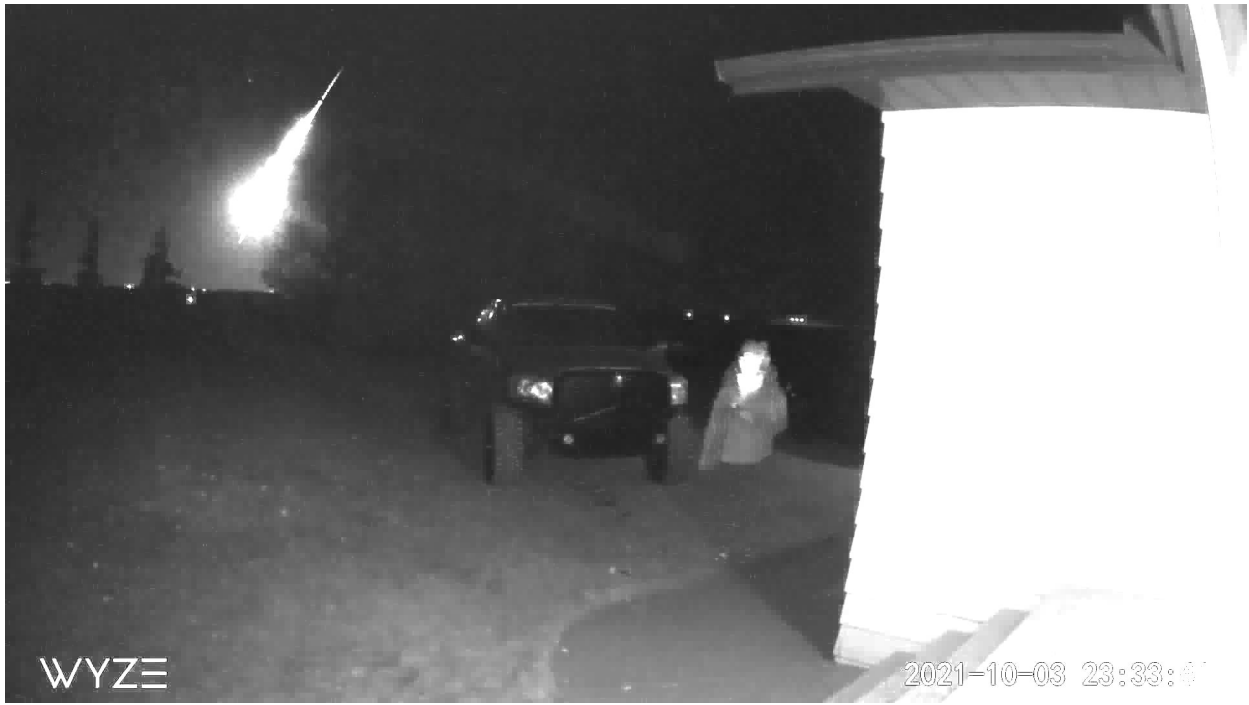

**Figure 5**. A maximum stack of all frames capturing the fireball looking West from Delacour, AB.

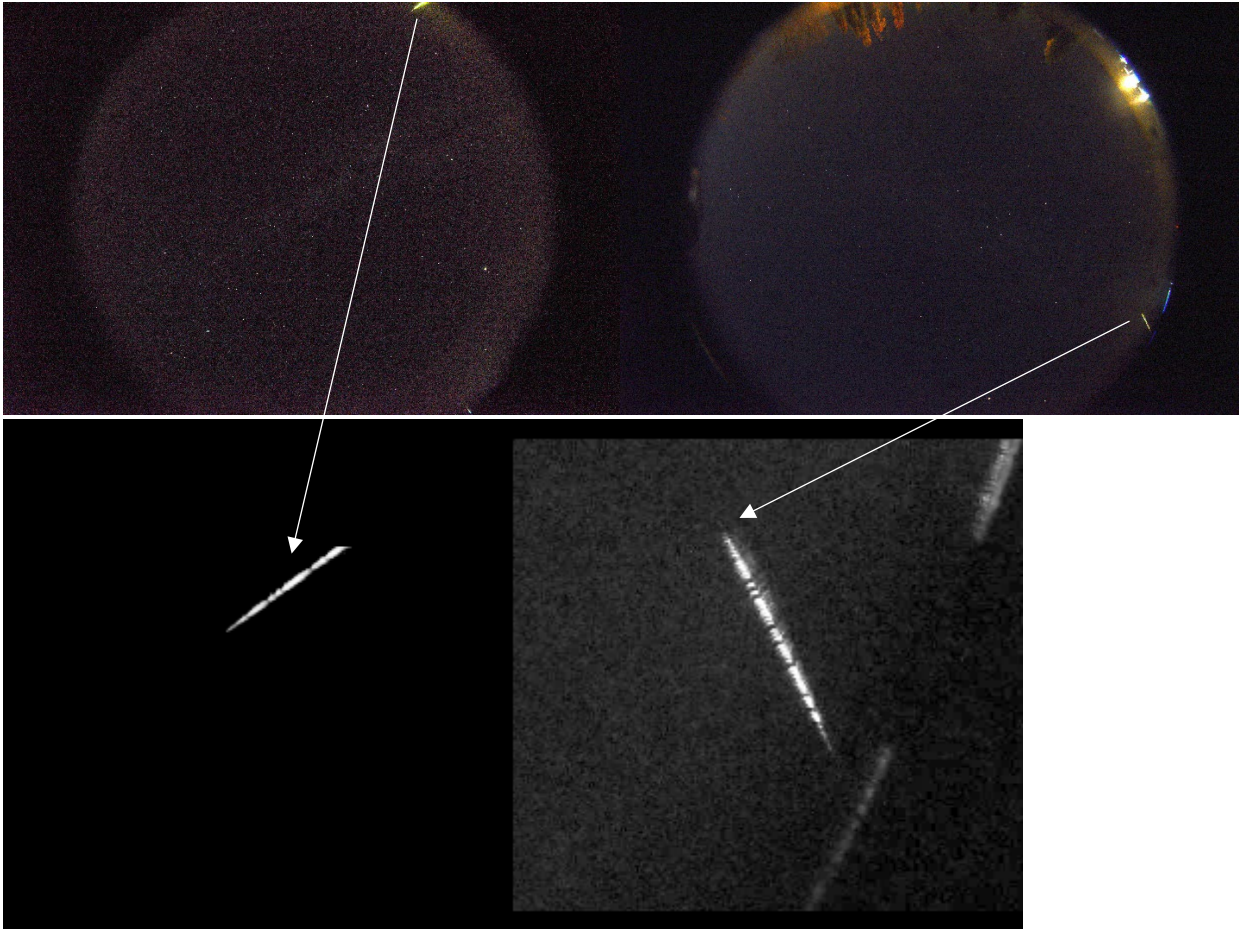

**Figure 6**. MORP 2.0 images of the Golden fireball as taken from Mattheis Ranch (top left) and Vermilion (top right), Alberta. Lower insets are enlargements of the fireball image showing the breaks created by the de Bruijn temporal sequencing of the LCD shutter (Howie et al., 2017).

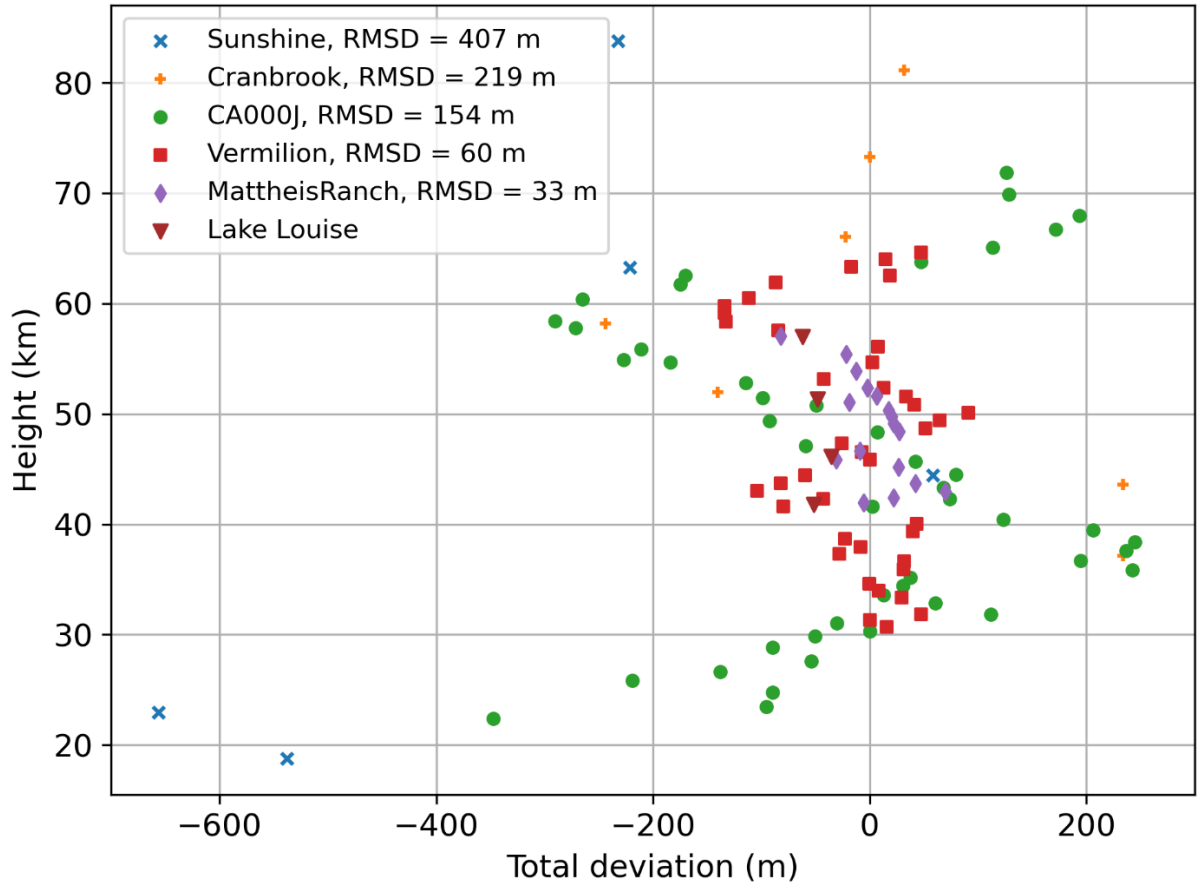

**Figure 7**. Spatial residuals from all cameras relative to the best-fit trajectory for the Golden fireball. The trajectory solution uses both relative timing and look angle per frame (or per break for GFO cameras) following the method described in Vida et al. (2020). Note that Lake Louise was not used in the solution as it lacked timing, but the look angles are draped on the master trajectory solution.

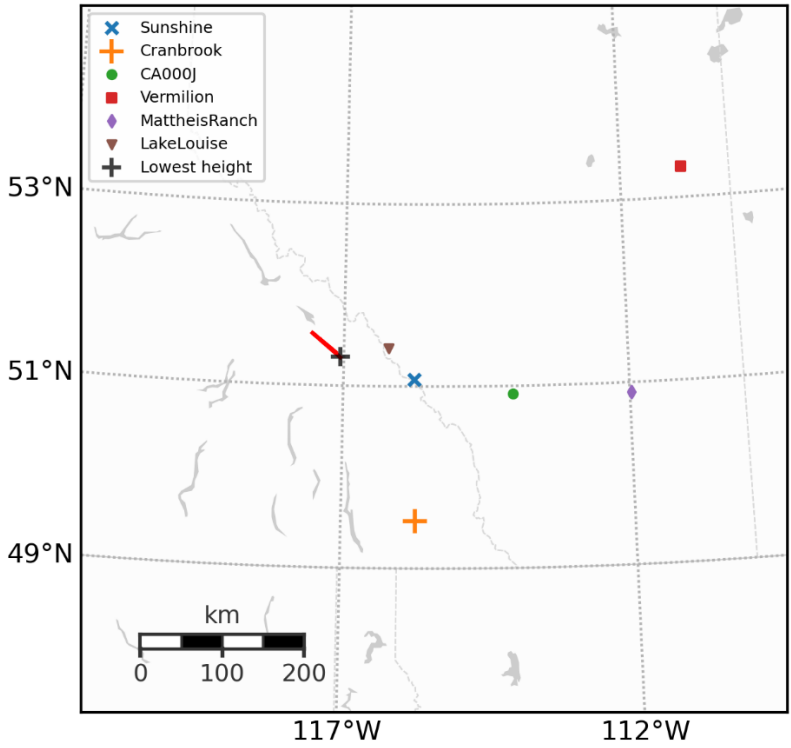

**Figure 8**. Fireball ground path (red line) and end point (black +) relative to camera/video stations used in astrometric solution as noted in the legend..

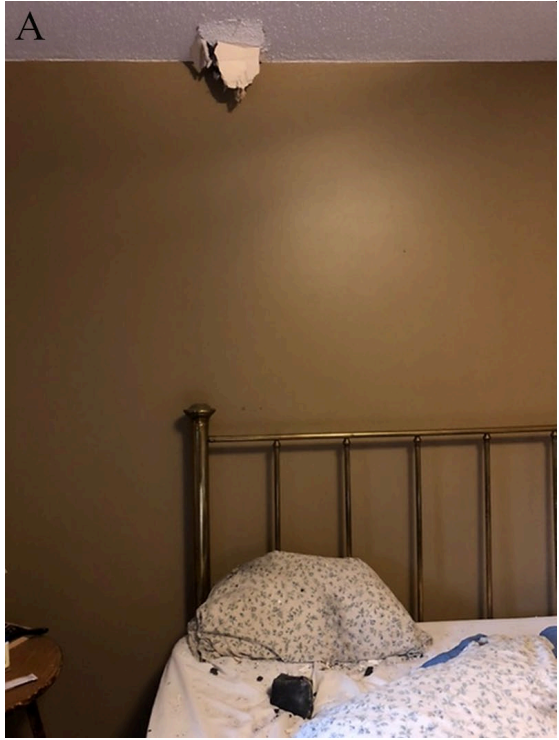
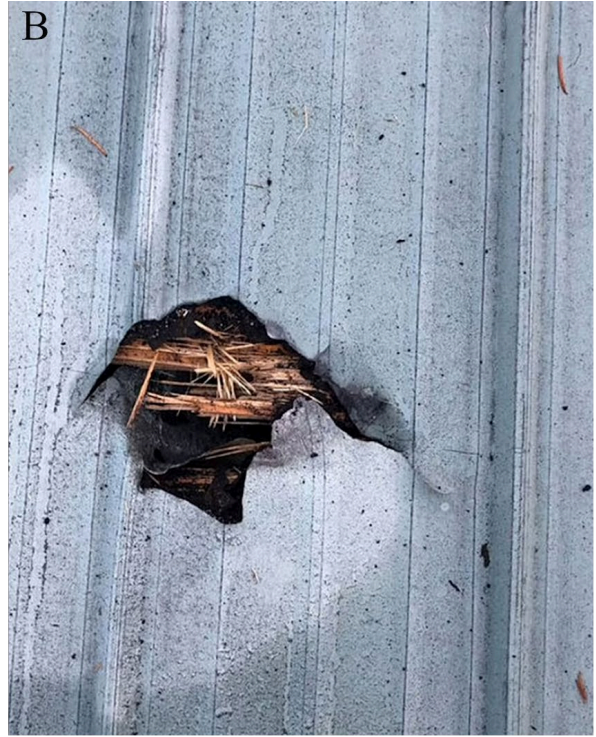
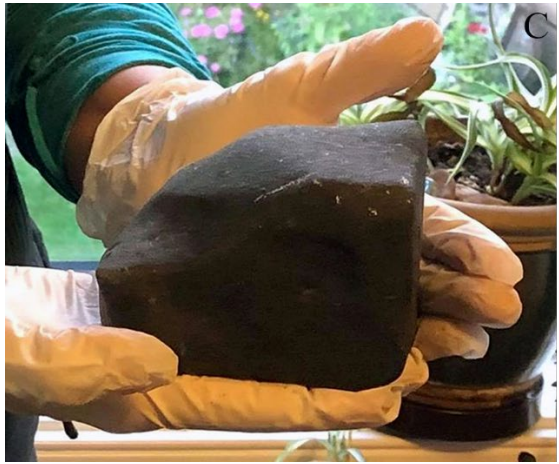
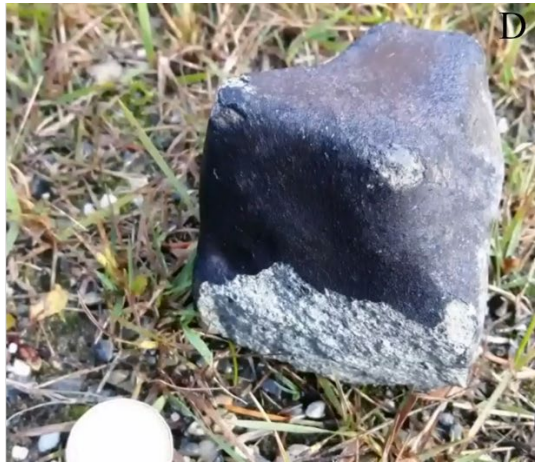

**Figure 9**. Arrival of the Golden meteorite. A) Meteorite as a complete individual fragment in situ on bed after falling through bedroom roof. Other dark debris on bed is roof asphalt and ceiling plaster. B) Hole in the tin, asphalt and wood roof caused by meteorite impact as seen from top of roof, top looking east. Width of roof hole is ~10 cm. C) Closeup of 1.3 kg Hamilton fragment which hit bed. D) The 0.9 kg "Calgary" fragment in find location on roadway, with mostly complete fusion crust and a broken surface showing a grey-green chondritic interior (scale coin is 2.6 cm diameter).

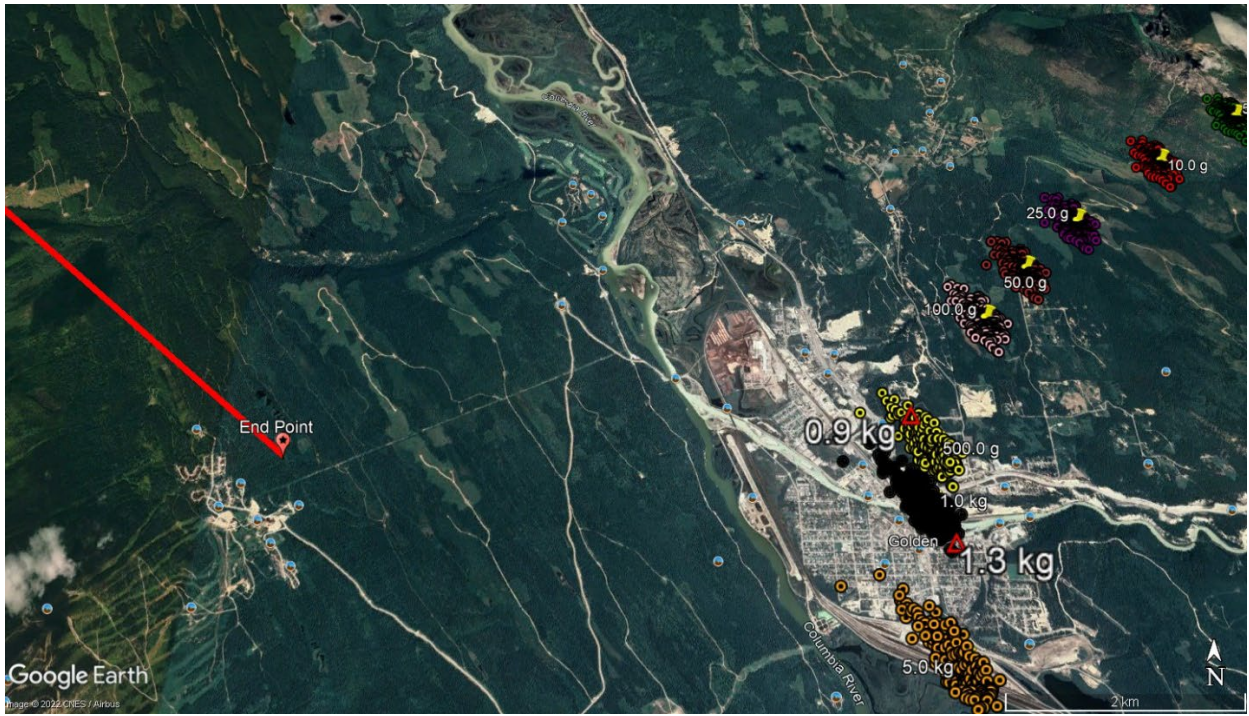

**Figure 10**. Darkflight model predictions for the ground footprint of selected model masses in Golden meteorite fall ellipse. The fireball trajectory and endpoint are shown as the red line. The individual meteorite recoveries are shown as large red triangles and their masses are given in large, bold white numbering. The range of Monte Carlo model fall zone locations for each model mass released at the endpoint as described in the text for model masses from 5 g (upper right) to 5 kg (lower center) is shown.

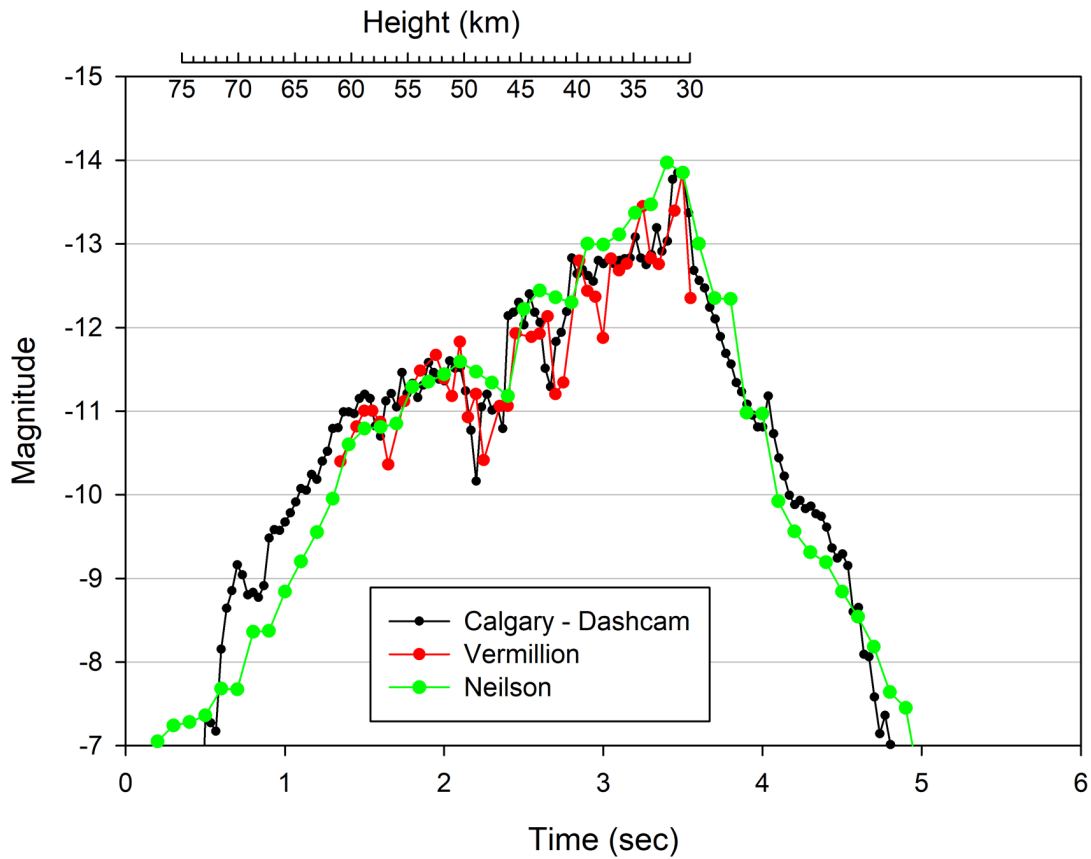

**Figure 11**. Lightcurve as a function of time for the Golden fireball from three cameras. Here the Vermilion camera is used as the absolute calibration in time and peak brightness and the other cameras are scaled to match. The height axis at top is constructed relative to the time (which is the independent variable) assuming a constant velocity. Below 30 km deceleration becomes significant and the simple linear height scaling no longer is accurate.

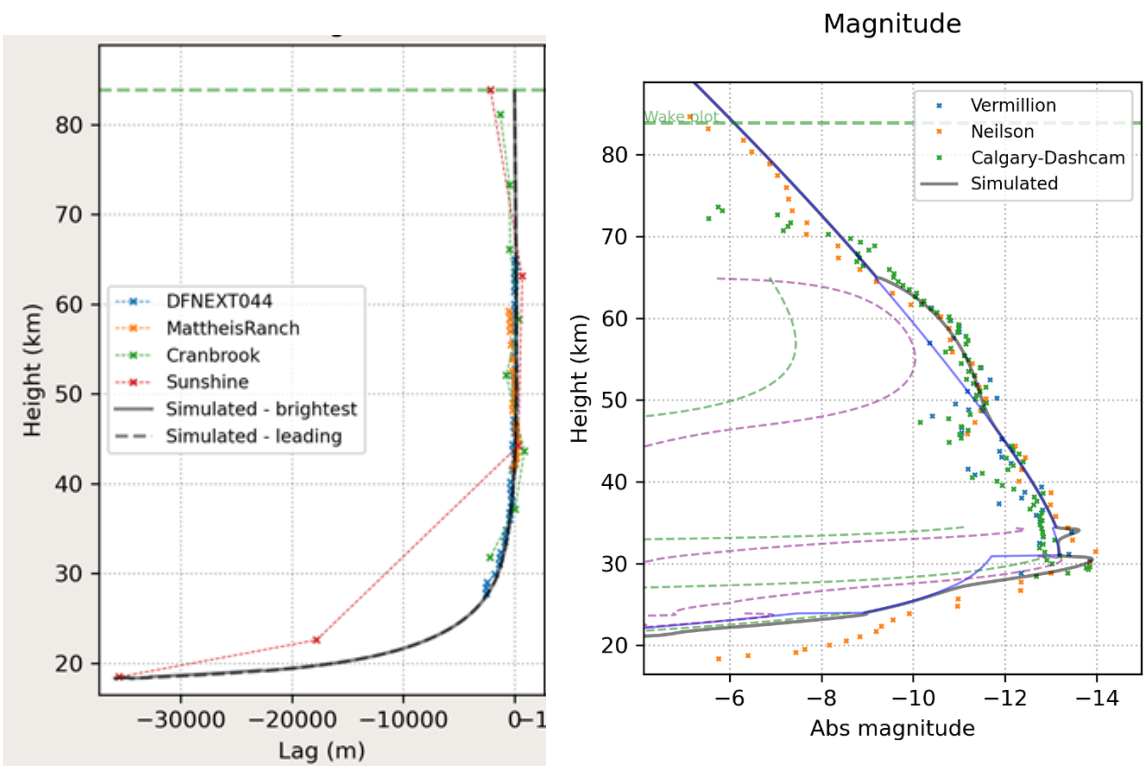

**Figure 12**. Semi-empirical ablation model of Borovička et al. (2020) applied to the Golden fireball and compared to observations. The left plot shows the apparent lag between the modelled leading fragment and the lags measured at each station. The right plot shows the simulated brightness from the semi-empirical model summing ablation of all fragments and dust at each time step as compared to the observed lightcurves (see Fig. 11). See Table 5 for fragment heights of release and other details of the fragmentation solution. Here DFNEXT044 is the Vermilion camera.

**Figure 13**. Infrasound record from I10CA (left) and I56US (right). The bolide signal is highlighted by the vertical green line and is centered around 0650 UTC and 0553-0555UT at each station respectively. Shown is the bandpassed waveform between 0.7 - 2 Hz and 0.5 – 8 Hz for I10 and I56 (bottom plot) in units of Pascals for element 1 of each array. Here we have used 120 and 60 sec windows with 50% and 90% overlap and found the best beam correlation (top plot), array trace velocity (second from top) and apparent backazimuth (third plot from top).

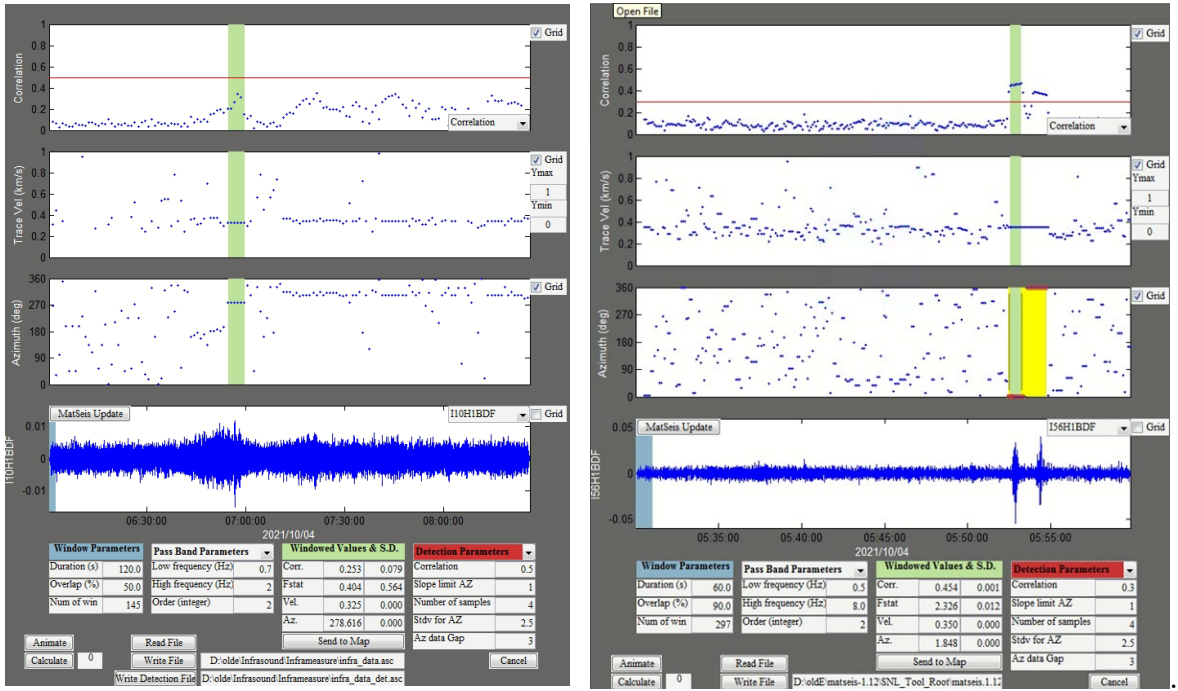

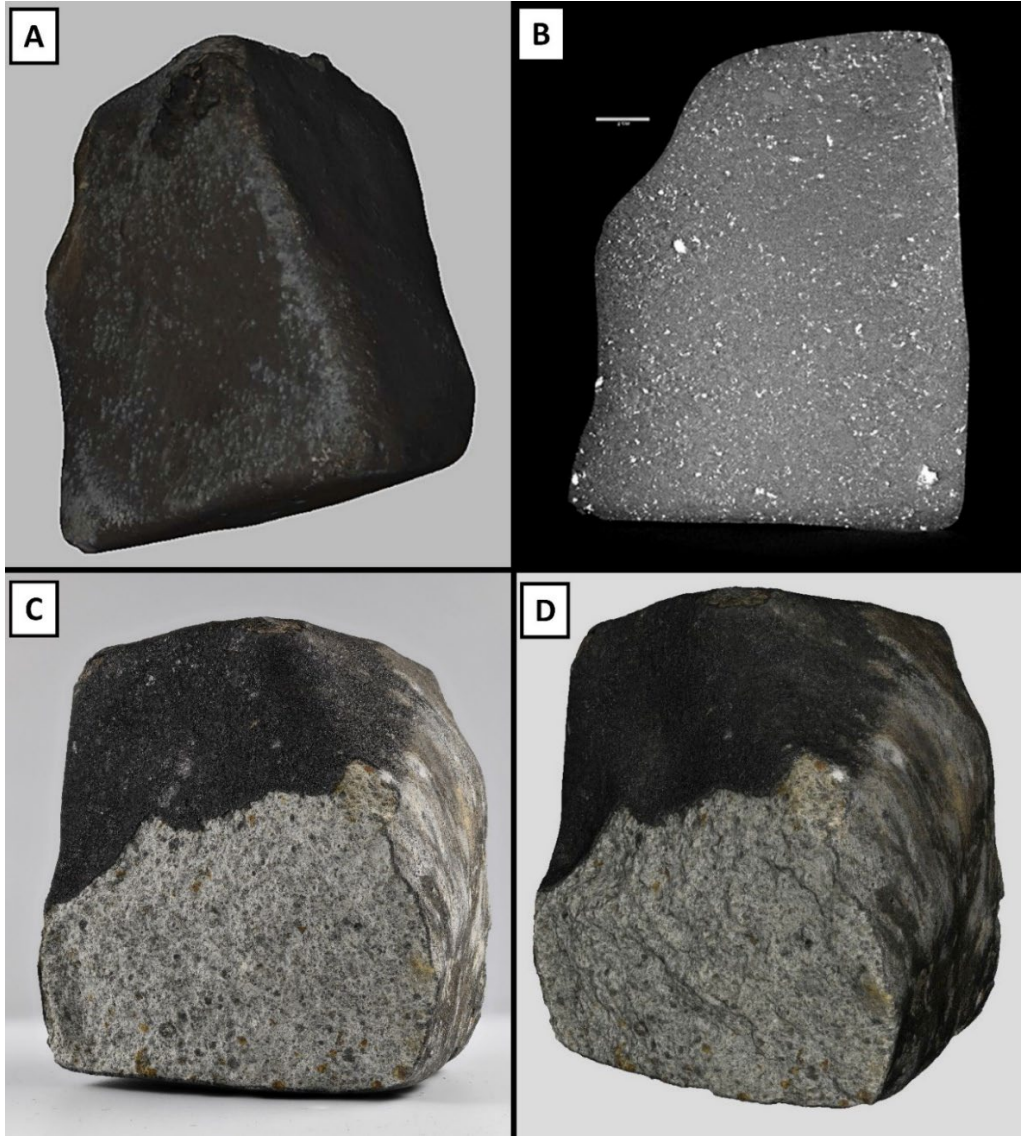

**Figure 14**. Model images of the Hamilton (main mass) 1.27 kg fragment and Calgary (0.92 kg) fragments of the Golden meteorite. A) Visible light 3D reconstruction of the Hamilton fragment, showing a faithful representation of surface albedo and colour features. Note the dark adhered asphalt lump at the top of the view. B) Virtual slice of X-ray CT reconstructed 3D model, corresponding to immediate subsurface of the main face showing in A). Scale bar is 1 cm. Bright objects in the slice are radiodense FeNi metal and sulfide, mostly disseminated as sub-mm grains, but with some larger metal-sulfide aggregates and an irregular elongate aggregate subparallel to the surface at upper right. C) A colour focus-stacked image of the Calgary fragment. Note that the interior had not noticeably started weathering with only a week's exposure to the terrestrial atmosphere when collected; the first rain had occurred just the morning of collection day, Oct. 10$^{th}$; although dried and stored in a nitrogen atmosphere within two days it had noticeably started weathering by imaging date Nov. 22$^{nd}$ presumably due to rain exposure on the day of collection. D) An example of the photogrammetric model of the Calgary fragment. Note the ground-scuffed side to the right and the light gray interior revealed by the broken surface on the fronting surface.

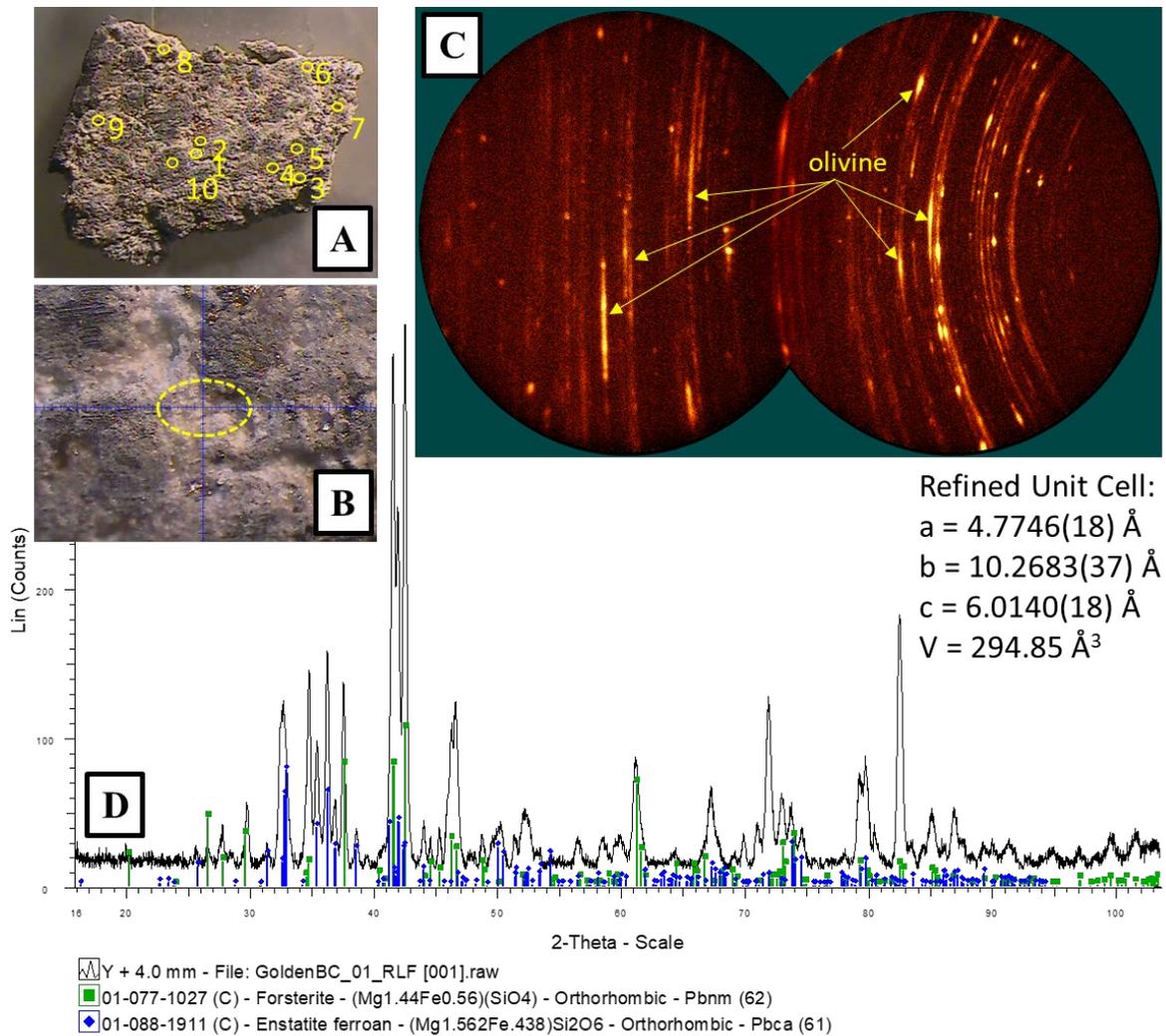

**Figure 15**. In situ micro X-ray diffraction analysis of olivine and pyroxene grains in Golden – Hamilton fragment. A) Target 300 μm in situ XRD beam spot locations on 0.7 g slice of Golden. Note dark chondrules in lighter, friable matrix with fusion crust at left edge. Width of view is 12 mm. B) Target camera colour image for location 1, with XRD beam spot footprint indicated by dashed ellipse having minor axis of 300 μm (Rupert et al., 2020). Target region consists mostly of matrix grains. C) 2D X-ray diffraction pattern of location 1 in two frames, with the undiffracted X-ray beam lying just to right of image and diffracted spots lying with increasing 2-theta angle to the left across image. Diffraction presents a mix of fine grains in Debye "powder" rings and coarser grains showing individual, slightly streaked diffraction spots reflecting mineral strain. Some spots and rings with 2-theta angles identified as forsteritic olivine are arrowed. D) Integrated 2-theta versus intensity plot of location 1, allowing peak positions to be searched with the ICDD database to identify and index forsterite olivine (green square sticks) and ferroan enstatite (blue diamond). Indexed peak positions for olivine were used to refine the unit cell dimensions and volume (V) in Angstroms given at right, which are used to estimate the Fe-Mg composition of olivine. See text for details.

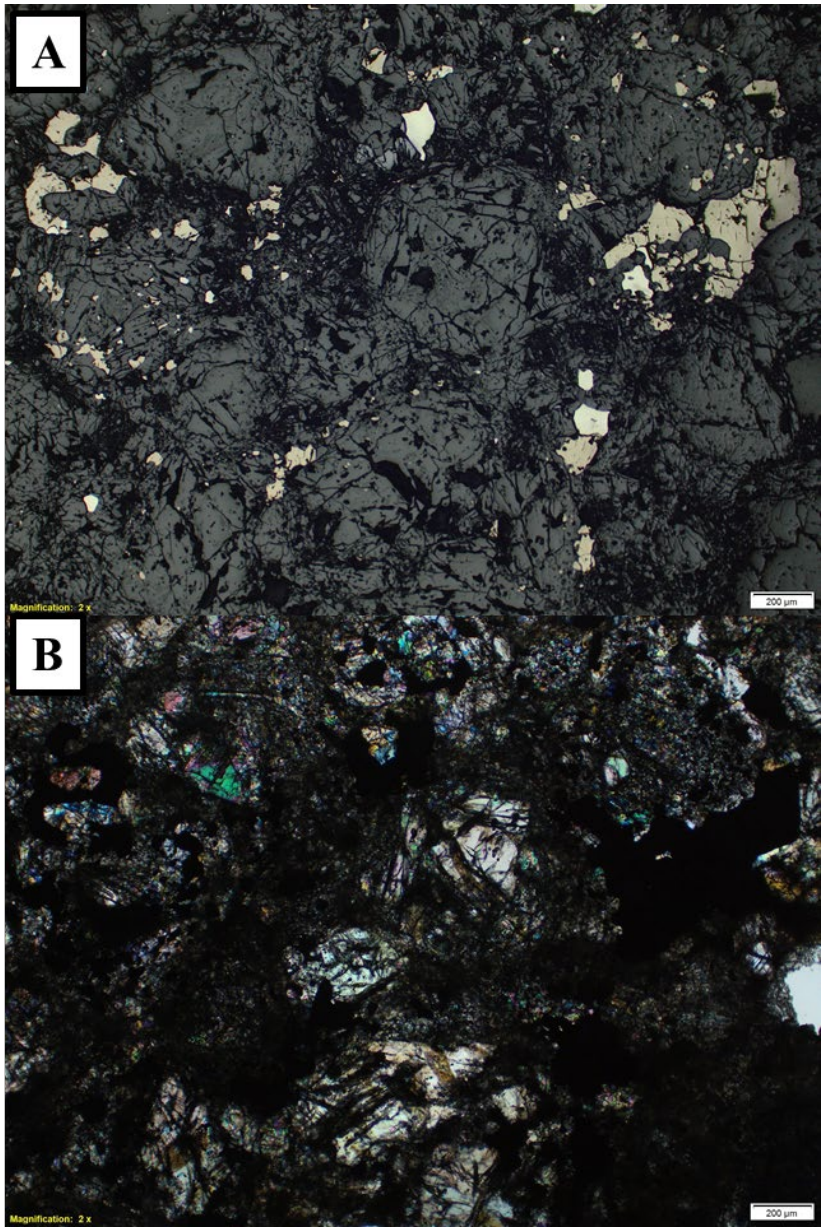

**Figure 16**. Polished thin section images of the Hamilton – Golden sample. Scale bar at lower right is 200 µm. A) Reflected light image showing subhedral metal (bright white) and sulfide (dimmer tan) grains and aggregates occurring mainly in matrix between poorly-defined silicate chondrules. Sulfide predominates over metal in this view. Darkest interstitial regions are pore space and fine matrix. B) Same view in cross-polarized transmitted light.

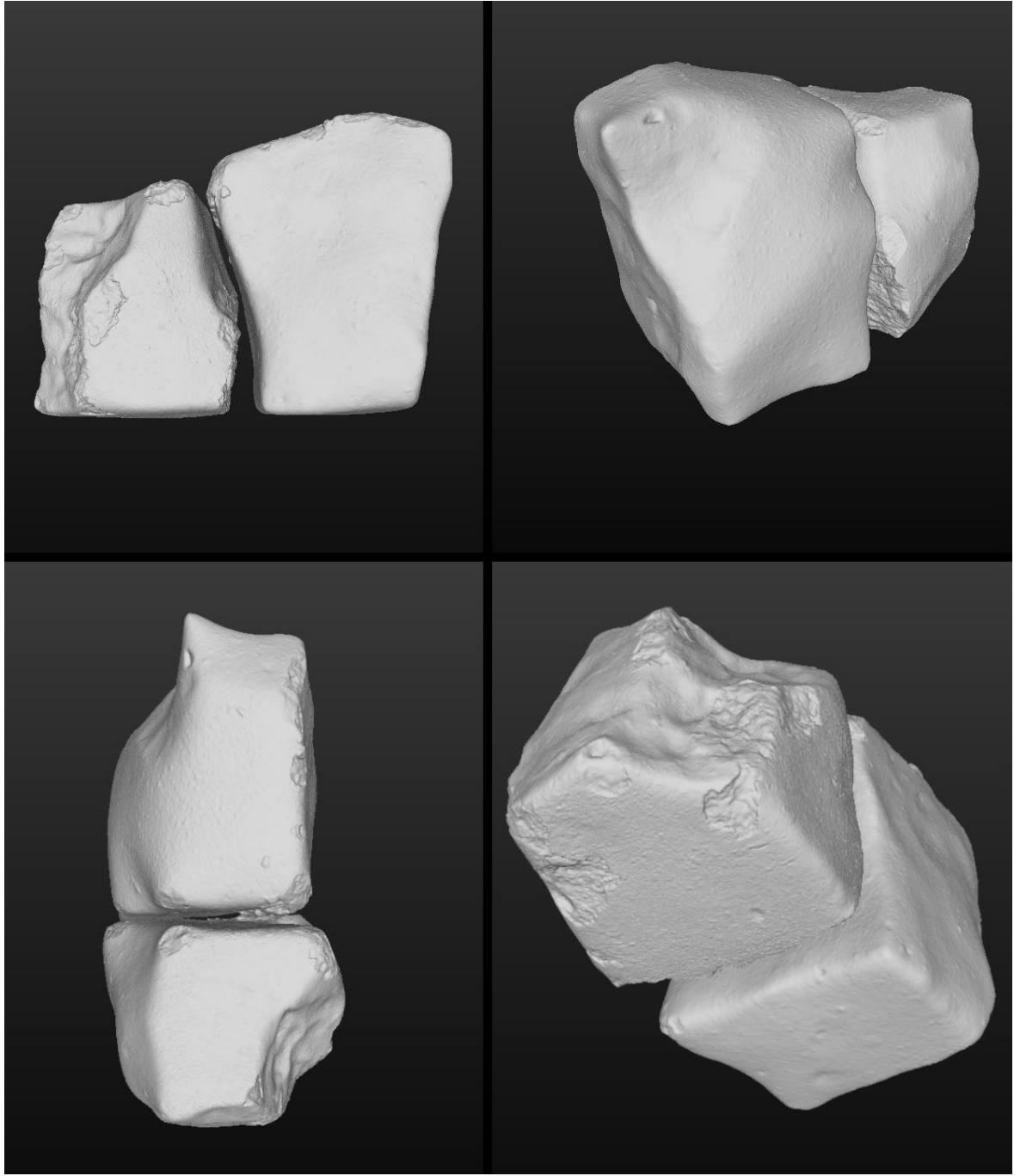

**Figure 17.** Four different views of a suggested possible pairing of the two recovered Golden meteorites shown in Meshlab software where the 3D models were manipulated. This orientation aligns the flat faces coplanar, has generally matching contours on the "shared" separation faces (note that this assumes that the largest broken surface on the Calgary stone is a thin spalled plate),

and roughly matches the widths/surfaces of the two "cubes" in all three dimensions. The largest side of the Hamilton fragment is 11cm for scale.

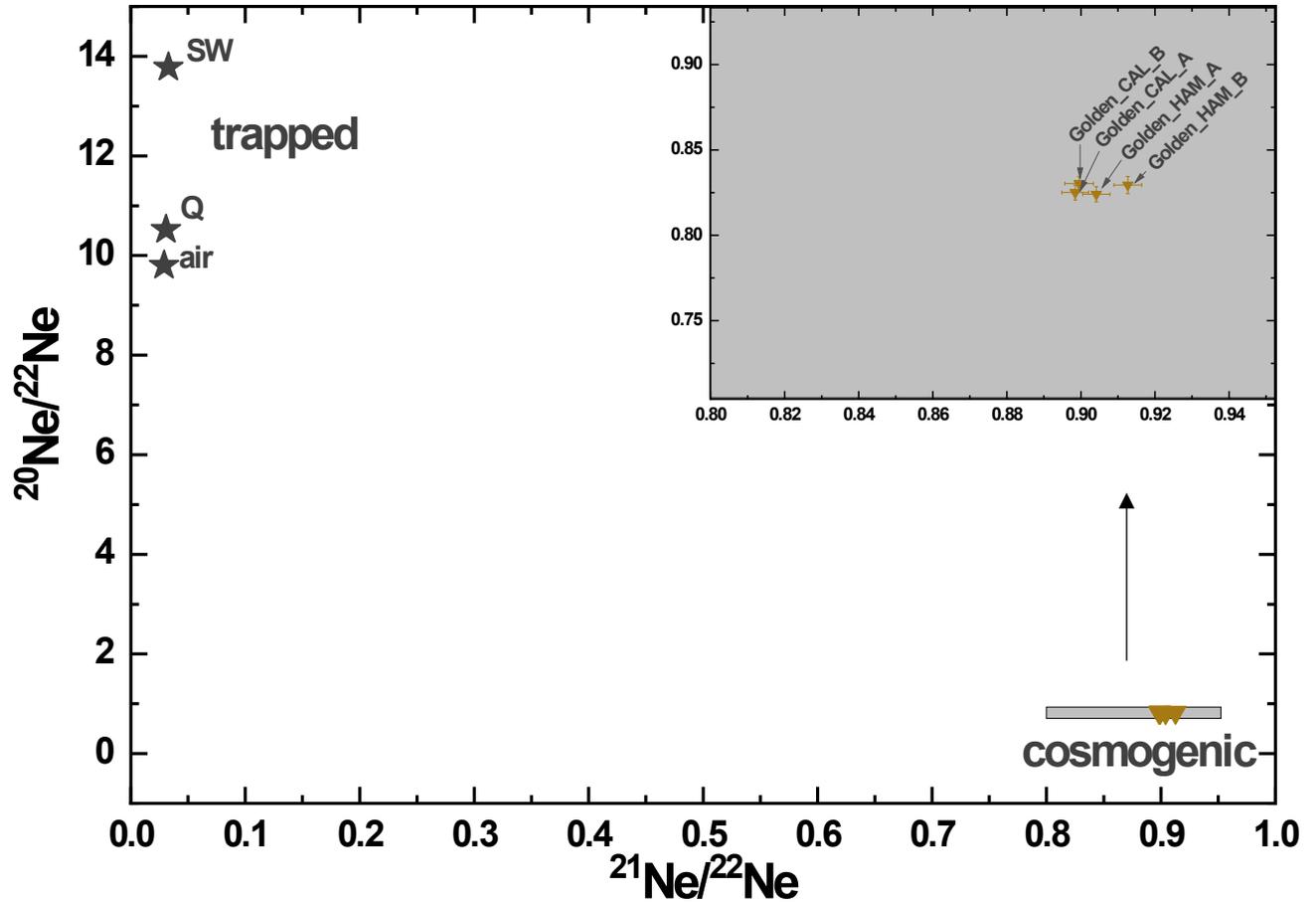

**Figure 18.** The Ne isotopic compositions of the four Golden samples analyzed is purely cosmogenic. The range for cosmogenic Ne of chondrites shown here is between 0.704–0.933 for $^{20}Ne/^{22}Ne$ and 0.800–0.952 for $^{21}Ne/^{22}Ne$ (Wieler 2002).

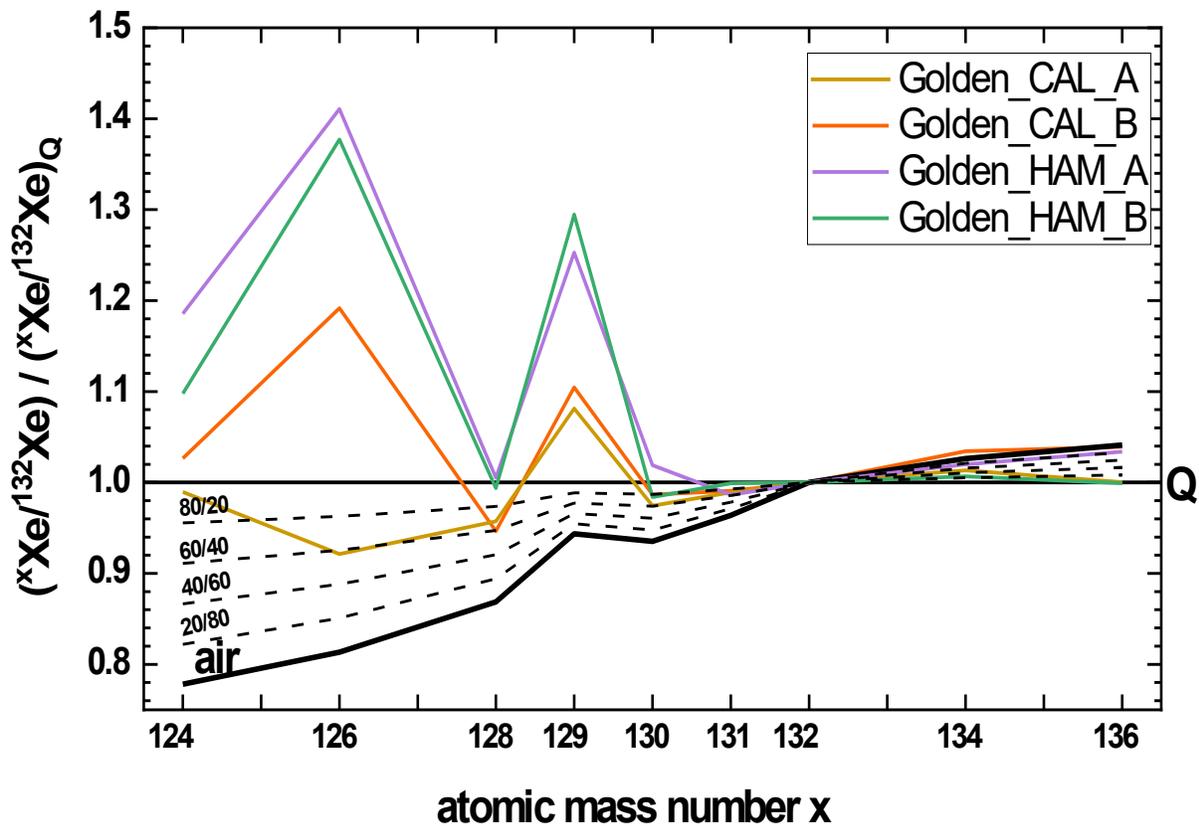

**Figure 19.** Xe isotopic compositions of the four aliquots and air normalized to Xe-Q from Busemann et al. (2000). The dashed lines give the isotopic compositions of Q mixed with air to varying ratios. While the Calgary aliquots show some contributions from air (Golden_CAL_A in e.g. $^{126,128}$Xe Golden_CAL_B in $^{134,136}$Xe and the much lower $^{129}$Xe/$^{132}$Xe ratios in both compared to Golden_HAM) those from Hamilton seem to be largely unaffected. Additionally, all four aliquots show some $^{129}$I-derived $^{129}$Xe excess.

**Figure 20.** Comparison of measured $^{10}$Be concentration in two samples of Golden L/LL chondrite with calculated $^{10}$Be depth profiles in L-chondrites with radii of 10-100 cm (A) and comparison of measured $^{10}$Be vs. $^{22}$Ne/$^{21}$Ne ratio in Golden with calculated relationship of $^{10}$Be vs. $^{22}$Ne/$^{21}$Ne (B). Both figures are consistent with a simple (one-stage) CRE age of >7 Ma at 5-10 cm depth in an object with a radius of 20-65 cm.

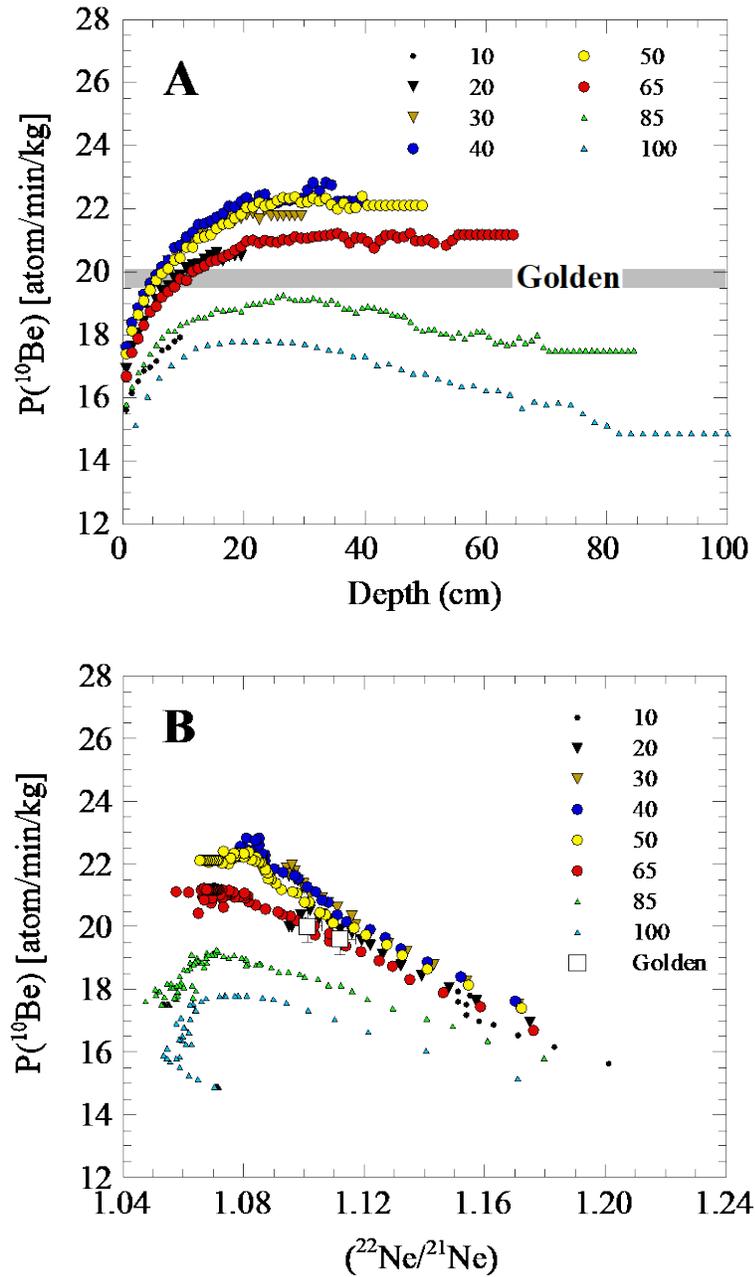

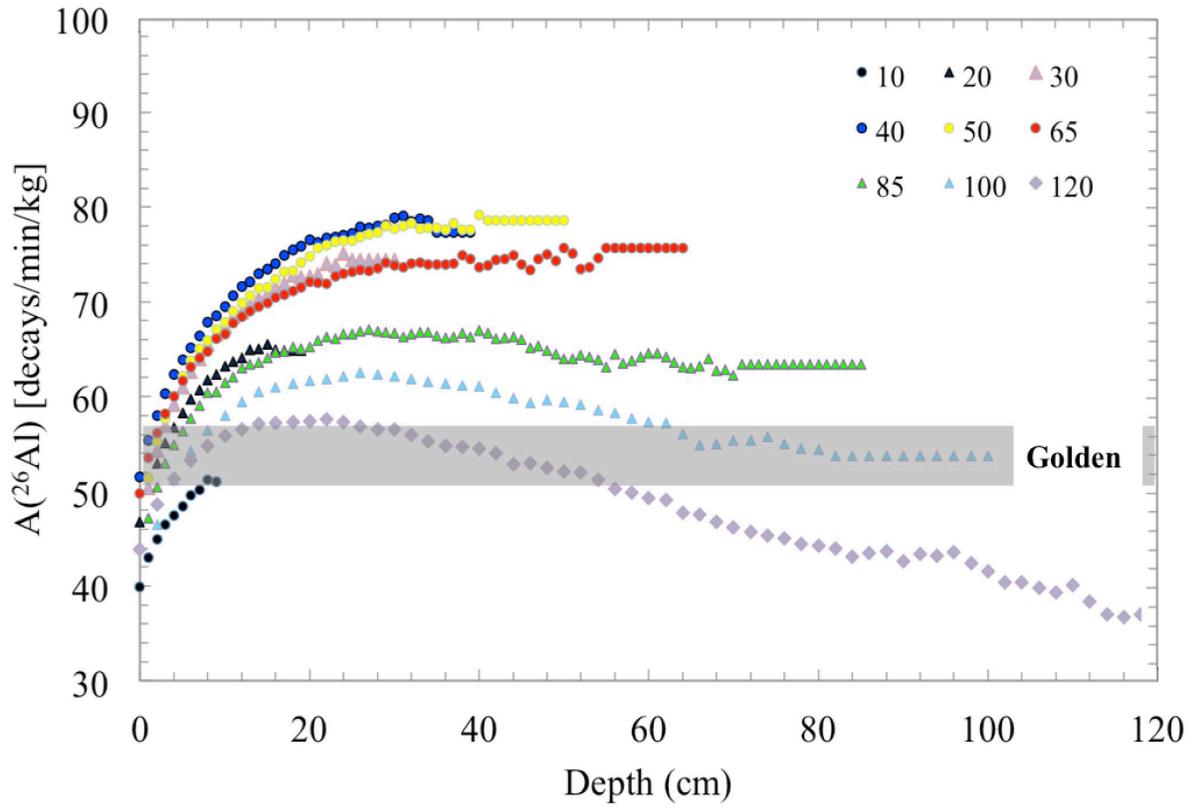

**Figure 21.** Comparison of measured $^{26}$Al concentration in the Hamilton sample of Golden with calculated $^{26}$Al depth profiles in L-chondrites with radii of 10-120 cm from Leya et al. (2009). The figure is consistent with 4-8 cm depth in an object with a radius of 10-120 cm.

**Supplementary Material**

**Appendix A: EPMA Results: Hamilton specimen**

**Appendix B: BSE elemental maps: Hamilton Specimen**

**Appendix C: Opaque chondrule: Calgary Specimen**